\documentstyle[prl,aps,epsf,multicol]{revtex}
\tighten
\renewcommand{\narrowtext} 
{\begin{multicols}{2}\global\columnwidth20.5pc} 
\renewcommand{\widetext}
{\end{multicols}\global\columnwidth42.5pc} 
\input psfig
\begin{document} 
\draft 
\title{Semiclassical theory of transport in a random magnetic field} 
\author{F. Evers, A. D. Mirlin,$^{*}$
D. G. Polyakov,$^{\dagger,\ddagger}$ and P. W\"olfle} 
\address{ Institut f\"ur Theorie der Kondensierten Materie,
Universit\"at Karlsruhe, 76128 Karlsruhe, Germany}
\maketitle

\begin{abstract} We present a systematic description of the
semiclassical kinetics of two-dimensional fermions in a smoothly
varying inhomogeneous magnetic field $B({\bf r})$. We show that the
nature of the transport depends crucially on both,
the strength of the
random component of $B({\bf r})$ and its mean value $\overline B$. 
For vanishing
$\overline B$, the governing parameter is $\alpha=d/R_0$, where $d$ is
the correlation length of disorder and $R_0$ is the Larmor radius in
the field $B_0$, the characteristic amplitude of the fluctuations of
$B({\bf r})$. While for $\alpha\ll 1$ the conventional Drude theory
applies, in the limit of strong disorder ($\alpha\gg 1$) most particles 
drift adiabatically along closed contours
and are localized within the adiabatic approximation. The
conductivity is then determined by the percolation of a special class of
trajectories, the ``snake states". The unbounded snake states
percolate by scattering at the saddle points of $B({\bf r})$ where the
adiabaticity of their motion breaks down. The external field
$\overline B$ is also shown to suppress
the stochastic diffusion by creating a
percolation network of drifting cyclotron orbits. This kind of
percolation is due only to the (exponentially weak) violation of the
adiabaticity of the rapid cyclotron rotation in the field $\overline
B$, leading to an exponentially fast drop of
the conductivity at large
$\overline B$. We argue that in the regime $\alpha\gg 1$ the
 crossover between the snake-state percolation and
the percolation of the drift orbits with increasing $\overline B$ is
very sharp and has the character of a phase transition (localization of
snake states) smeared exponentially weakly by non-adiabatic effects. 
The ac conductivity also reflects the dynamical properties of particles
moving on the fractal percolation network. In particular, we
demonstrate that the conductivity has
a sharp kink at zero frequency and falls off
exponentially  at higher frequencies. 
We also discuss the nature of the
quantum magnetooscillations. We report detailed numerical studies of
the transport in the field $B({\bf r})$: the results of the numerical
simulations confirm the analytical findings. 
The shape of the magnetoresistivity at $\alpha\sim 1$ found
numerically is in good agreement with the experimental data in the
fractional quantum Hall regime for the
vicinity of half-filling of the lowest Landau level.
\end{abstract}

\narrowtext

\section{Introduction}
\label{s1}

The transport properties of two-dimensional ($2d$) particles moving in
a spatially random magnetic field (RMF) ${\bf B}({\bf r})$ oriented
perpendicularly to the plane have attracted considerable interest in
the last few years. This interest is largely motivated by the
relevance of the problem to the composite-fermion (CF) description
\cite{jain,halperin93} of a half-filled Landau level. Within this approach,
the electron liquid in a strong magnetic field is mapped -- by means
of a Chern-Simons gauge transformation -- to a fermion gas subject to
a weak effective magnetic field. Precisely at half-filling, the
average value of the Chern-Simons gauge field compensates the
effect of the external magnetic field. The RMF appears in this model
after taking static disorder into account: fluctuations of the local
filling factor induced by the random potential of impurities lead to a
local mismatch between the gauge and external magnetic fields. A
number of observations \cite{willett97} of Fermi-surface features near
half-filling give strong experimental support to the model of the
effective magnetic field. Apart from the composite-particle models
involving fictitious fields, $2d$ electron systems with a real RMF can
be directly realized in semiconductor heterostructures by attaching to
the latter superconducting \cite{bending90,geim92} or ferromagnetic
\cite{ye96,zielinski98} overlayers.

The peculiarity of transport properties of $2d$ electrons in a random
field $B({\bf r})$ shows up most distinctly in systems with {\it
smooth} inhomogeneities. The case of long-range disorder is most
important also experimentally -- since the compressible state in a
half-filled Landau level is observed in high-mobility samples. In the
latter, a large correlation radius of potential fluctuations, $d$, is
determined by a wide ``spacer'' between the electron gas and the doped layer
containing ionized impurities. Likewise, inhomogeneities of the RMF
created by the ferromagnetic overlayers in \cite{ye96,zielinski98}
appear to be fairly long-ranged. The large value of the correlation
radius $d$ (as compared to the interelectron distance) allows to
describe the electron kinetics {\it quasiclassically}. 

It is well known that quantum interference effects may cause
localization of noninteracting particles in an infinite 2d system even
for arbitrarily weak disorder. This has been shown to be the case for
charged particles in a random magnetic field \cite{amw94,kim95}.
Specifically, the random magnetic field
problem belongs to the unitary universality class, with
the localization length $\xi$ growing extremely fast with the dimensionless
conductance $g=\sigma_{xx}/(e^2/h)$,  
\begin{equation}
\label{e0}
\xi\propto\exp(\pi^2 g^2).
\end{equation}
These  theoretical results are  in full agreement with the
recent extensive numerical study \cite{furusaki98}. 
According to (\ref{e0}), already for $g\gtrsim 1.5$ the localization length is
larger than any reasonable system size, and the quasiclassical approach is
fully justified.  

Let us stress that we consider the situation, in which the smooth
random magnetic field constitutes the only 
type of disorder present. This should be contrasted with the
starting point of Refs.~\cite{smith94}, where the main contribution to
the resistivity was assumed to be
given by a short-range random scalar potential (treated
within the relaxation time approximation for the collision integral in
the Boltzmann equation), while a weak long-range correlated random
magnetic field was considered as a small perturbation.

The purpose of this paper is to examine the transport in a long-range
RMF in detail, with particular emphasis on the conductivity in an
external (homogeneous) magnetic field ${\overline B}
=\langle B({\bf r})\rangle$ and/or at finite
frequency $\omega$. As we will see, the physics of the problem depends
crucially on the values of $\overline B$ and $\omega$, leading
to a rich variety of different transport regimes. We
complement the analytical analysis by numerical simulations. The
importance of the latter is due to the fact that in the most
interesting part of the parameter space the transport is dominated by
the phenomenon of percolation, so that only estimates ``by order of
magnitude" are available at the analytical level.

In Sec.\ \ref{s2} we study the dc conductivity in a long-range RMF
$\delta B({\bf r})$ at zero $\overline B$. The character of the
${\overline B}=0$ transport is determined by the parameter
$\alpha=d/R_0$, where $R_0=v_F(mc/eB_0)$ is the Larmor radius in the
field $B_0$ which is a characteristic amplitude of the fluctuations,
$v_F$ the Fermi velocity. At $\alpha\ll 1$, the classical dynamics is
of conventional diffusive nature, and the conductivity can be
calculated in the Born approximation, which gives $\sigma_{xx}\sim
(e^2nd/mv_F)/\alpha^2$, where $n$ is the particle density. With
increasing $\alpha$, a crossover to the percolating kinetics takes
place, and at $\alpha\gg 1$ the conductivity is determined by a small
{\it fraction} of classical trajectories -- so-called ``snake states''
\cite{chklovskii93,lee94}. The extended snake states percolate through
the strongly disordered system by winding around the lines of zero
$B({\bf r})$ and yield $\sigma_{xx}\sim (e^2nd/mv_F)/\alpha^{1/2}{\cal
L}$, which falls off slowly with increasing strength of disorder [here
${\cal L}=(\ln\alpha)^{1/4}$ is a weakly varying logarithmic
factor]. The crossover from the $1/\alpha^2$ to $1/\alpha^{1/2}$
behavior of $\sigma_{xx}$ at $\alpha\sim 1$ is confirmed by our
numerical simulations. Furthermore, the latter allow us to find the
numerical value of $\sigma_{xx}$ for the CF problem (for which
$\alpha\simeq 1/\sqrt{2}$ lies in the crossover region).

In Sec.\ \ref{s3} we consider the case of strong $\overline B$. The
transport through the snake states, examined in the previous section,
is a dominant mechanism of the conductivity at strong disorder and
{\it zero} mean $\overline B$. Increasing $\overline B$ also leads to
a suppression of the conventional diffusive motion and a transition to a
percolation regime, even if $\alpha\ll 1$. The physics of this
phenomenon is, however, quite distinct from the snake-state
percolation. In the limit of large $\overline B$, the trajectories 
are of the form of drifting cyclotron orbits, and the dynamics
is governed by an adiabatic invariant (magnetic flux through one
cyclotron orbit). In
the adiabatic approximation, the particles thus drift along the closed
magnetic field contours and hence are localized. It is only the weak
{\it nonadiabatic} scattering between drift trajectories that yields a
finite conductivity. This localization effect is similar to the
formation of a ``stochastic web" \cite{fogler97,zaslavsky91} in a
slowly varying scalar random potential in the presence of an external
homogeneous magnetic field. The conductivity due to the weak
nonadiabaticity falls off {\it exponentially} fast with increasing
$\overline B$. To calculate the function $\sigma_{xx}({\overline B})$
explicitly, one should consider the averaging procedure with care. An
essential point is that the nonadiabatic transitions that determine
the conductivity occur in {\it rare} fluctuations distributed sparsely
along the percolating trajectories. These rare spots are characterized
by anomalously sharp changes of the RMF. Using the optimum-fluctuation
method, we find a Gaussian behavior of the conductivity in the limit
of large $\overline B$: $\ln\sigma_{xx}=-A(\alpha)\left({\overline
B}/B_0\right)^2$, where the coefficient $A(\alpha)$ scales as
$\alpha^{4/3}$. Note that the conductivity in the CF problem ($A\sim
1$) falls off {\it sharply} beyond a {\it small} deviation from
half-filling. It is worthwhile to notice also that the exponential
falloff of $\sigma_{xx}$ with increasing $\overline B$ is limited by
inelastic scattering and by scattering on short-range static
inhomogeneities -- these scattering processes allow an electron to
change drift trajectories and thus provide a mechanism of percolation
competing with the nonadiabatic transitions
\cite{isichenko92,zielinski98}.

While increasing $\overline B$ drives the system into the adiabatic
regime at any $\alpha$, the manner in which the conductivity crosses
over into this regime is qualitatively different in the cases of weak
and strong disorder. The situation is especially interesting at
$\alpha\gg 1$, where the transport regimes controlled by the snake
states (weak $\overline B$) and by the nonadiabatic dynamics of the
cyclotron orbits (strong $\overline B$) are separated by a sharp {\it
transition} accompanied by an abrupt change of $\sigma_{xx}$. The
transition occurs at ${\overline B}\sim B_0/\alpha^{2/3}$, where
$\overline B$ becomes larger than the characteristic amplitude of the
magnetic field at the nodes of the conducting network. At the critical
point, the percolation network formed by the extended snake states
falls apart into disconnected clusters, while the nonadiabatic
scattering on the high-$\overline B$ side of the transition is
strongly suppressed and yields only a slight smearing of the critical
singularity.

For $\alpha\sim 1$, which is the regime relevant to the CF problem, we
perform a numerical simulation to calculate the magnetoresistance
$\rho_{xx}({\overline B})$. The obtained curves are in good agreement
with experimental findings: they show a weak positive
magnetoresistance at low ${\overline B}$, crossing over to a falloff
of $\rho_{xx}$ with increasing ${\overline B}$ (in accordance with the
exponential suppression of $\sigma_{xx}$ in the adiabatic
regime). Further, we analyze the quantum oscillations of the
resistivity and show that, in contrast to the conventional
Shubnikov-de Haas effect in a short-range random potential, they
start to develop only when the dimensionless conductivity
$\sigma_{xx}/(e^2/h)$ drops down to a value of order of unity.

In Sec.\ \ref{s4} we discuss the transport in the RMF at finite
frequency $\omega$. We find strong deviations of the {\it ac}
conductivity from the Drude behavior, especially in the percolation
regime, i.e. when $\alpha$ and/or ${\overline B}$ is large.
At small $\omega$, we find a {\it nonanalytical}
$(\propto |\omega|$) contribution to the ac conductivity, which is
determined by returns of the particle to the same spatial regions
after a time $\sim 1/|\omega|$ (and is analogous to the long-time
``tails'' found in the Lorentz gas many years ago \cite{tails}). 
In the large-$\overline B$ regime and at higher frequencies, 
the ac conductivity takes
the form $\sigma_{xx}(\omega)\propto |\omega|^{3/7}$, since $\omega$
itself starts to determine the width of the percolating ``stochastic
web'' responsible for the conductivity. By contrast, in the
snake-state percolation regime ($\alpha\gg 1$, ${\overline B}=0$)
$\sigma_{xx}(\omega)$ shows only a weak
dispersion in the corresponding frequency range. 
At still larger frequencies the {\it ac}
conductivity starts to drop exponentially reflecting the ``ballistic''
motion of drifting orbits (or snake states) on short scales.
These analytical estimates are in agreement with our numerical simulations.

Sec.\ \ref{s5} summarizes our findings. The analytical results of
Sections \ref{s2}, \ref{s3} were partly presented in the Letter
\cite{mirlin98}.

\section{dc transport in zero mean magnetic field}
\label{s2}

We begin by formulating the model to be studied. We consider
non-interacting particles in the RMF $\overline{B}+\delta B({\bf r})$
with mean $\overline{B}$ and the correlator $\left<\delta B(0)\delta
B({\bf r})\right>=B_0^2f(r)$, where $f(0)=1$. We assume that the
function $f(r)$ is characterized by a single spatial scale, which is
the correlation length of the RMF. In particular, in the CF model with
the electron density $n$ equal to the charged impurity density $n_i$
we have
$B_0=(\hbar c/e)(k_F/\sqrt{2}d)$ and $f(r)=(1+r^2/4d^2)^{-3/2}$, where
$k_F^2=4\pi n$ (note that the electron gas is fully spin polarized
near $\nu=1/2$; we discard the spin degree of freedom throughout the
paper). 
In this section we confine ourselves to the case of zero $\overline
B$. The RMF with zero mean is characterized by two length scales: $d$
and the cyclotron radius $R_0$ in the field $B_0$. Defining the
parameter $\alpha = d/R_0$, we can distinguish the weak-RMF regime
$\alpha \ll 1$, where the mean free path $l\gg R_0\gg d$, and the
regime of strong fluctuations $\alpha\gg 1$, where one should expect
drastic deviations from the Drude picture. We will explore these two
limiting cases analytically. However, since the value of $\alpha$
corresponding to the CF problem lies in the crossover region, we will
turn to numerical simulations in order to get $\sigma_{xx}$ of the
CFs.

\subsection{Weak disorder}
\label{ss2a}

We start with the simple case of $\alpha\ll 1$. In this limit, the CF
trajectories are only slightly bent on the scale of $d$, so that the
Born approximation is valid. Accordingly, for the transport scattering
time one gets \cite{halperin93,aronov95} $\tau_{tr}^{-1}
=v_F^{-1}(eB_0/mc)^2 \int_0^\infty dr f(r) = 2\alpha^2v_F/d$, where
the CF effective mass $m=\hbar k_F/v_F$ is introduced. The Drude
conductivity at zero $\overline B$, $\sigma_{xx}=e^2n\tau_{tr}/m$,
then reads 
\begin{equation}
\label{e1} 
\sigma_{xx}={e^2\over h}{k_Fd\over 4\alpha^2},\quad\alpha\ll 1~.  
\end{equation}

\subsection{Dynamics of the snake states}
\label{ss2b}

\begin{figure}
\centerline{\epsfxsize=80mm\epsfbox{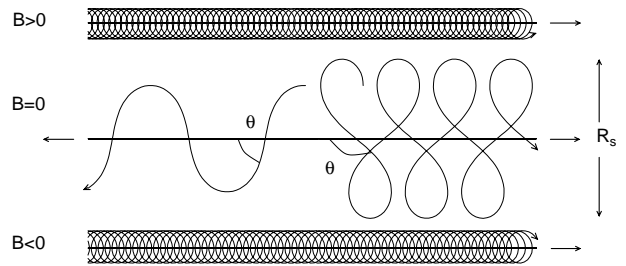}}
\vspace{3mm}
\caption{Types of trajectories in a strong random magnetic field:
drifting orbits along non-zero $B$ contours and
snake states near $B=0$ lines. Geometry of a snake state is
characterized by the angle $\theta$ ($0<\theta<\pi$) at which the
trajectory crosses the zero-field contour. Note that the direction of
motion of the snake state with $\theta<\theta_c$ (left) is opposite to
that for $\theta>\theta_c$ (right), where $\theta_c\simeq
131^\circ$. The width $R_s$ of the bundle of snake state trajectories
is also indicated.}
\label{fig1} 
\end{figure}

Let us now turn to the strong-RMF regime, $\alpha\gg 1$, keeping
$\overline{B}=0$. The seemingly innocent assumption about the chaotic
character of the particle dynamics, which enabled us to represent the
conductivity in the form $e^2n\tau_{tr}/m$, is not valid anymore. Most
particles are now out of play since they are caught in cyclotron
orbits drifting along the closed lines of constant $B({\bf r})$ (``van
Alfv\'en drift"). In the adiabatic limit, their drift trajectories are
periodic and so do not contribute to the conductivity. Still, however
large $B_0$ is, there are classical paths which are not localized and
percolate through the system by meandering around the lines of zero
$B({\bf r})$. The conductivity is determined by the particles that
move along these extended ``snake states" \cite{chklovskii93}
(Fig.~\ref{fig1}).

Note that there is only one single percolating path on the
manifold of the $B({\bf r})=0$ contours; yet, the conductivity is
nonzero since the snake-state trajectories form a bundle of finite
width, $R_s\sim d/\alpha^{1/2}$ (see Fig.~\ref{fig1}). 
The conducting network is made up of
those snake states that can cross over from one critical zero-$B$ line
to another.  This coupling of two adjacent percolating clusters occurs
near the critical saddle points of $B({\bf r})$, which are nodes of
the transport network. The crucial role of the saddle points is that
they break down the adiabaticity of the snake-state dynamics, as we
are going to explain below.

\begin{figure}
\centerline{\epsfxsize=82mm\epsfbox{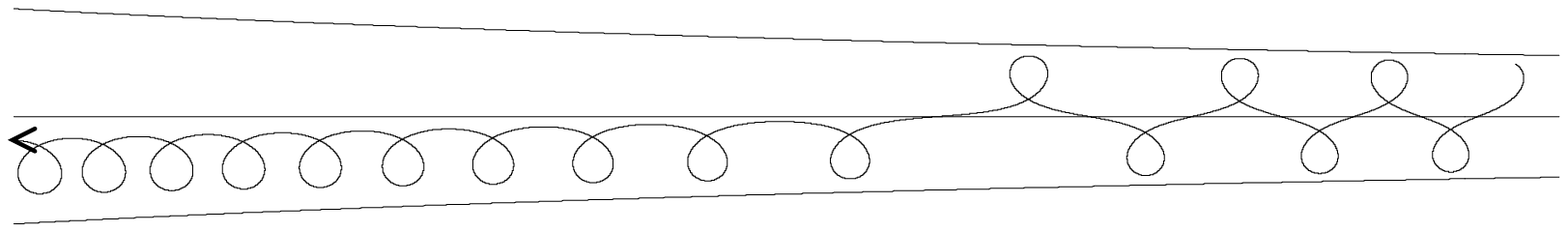}\vspace*{3mm}}
\vspace{5mm} 
\centerline{\epsfxsize=72mm\epsfbox{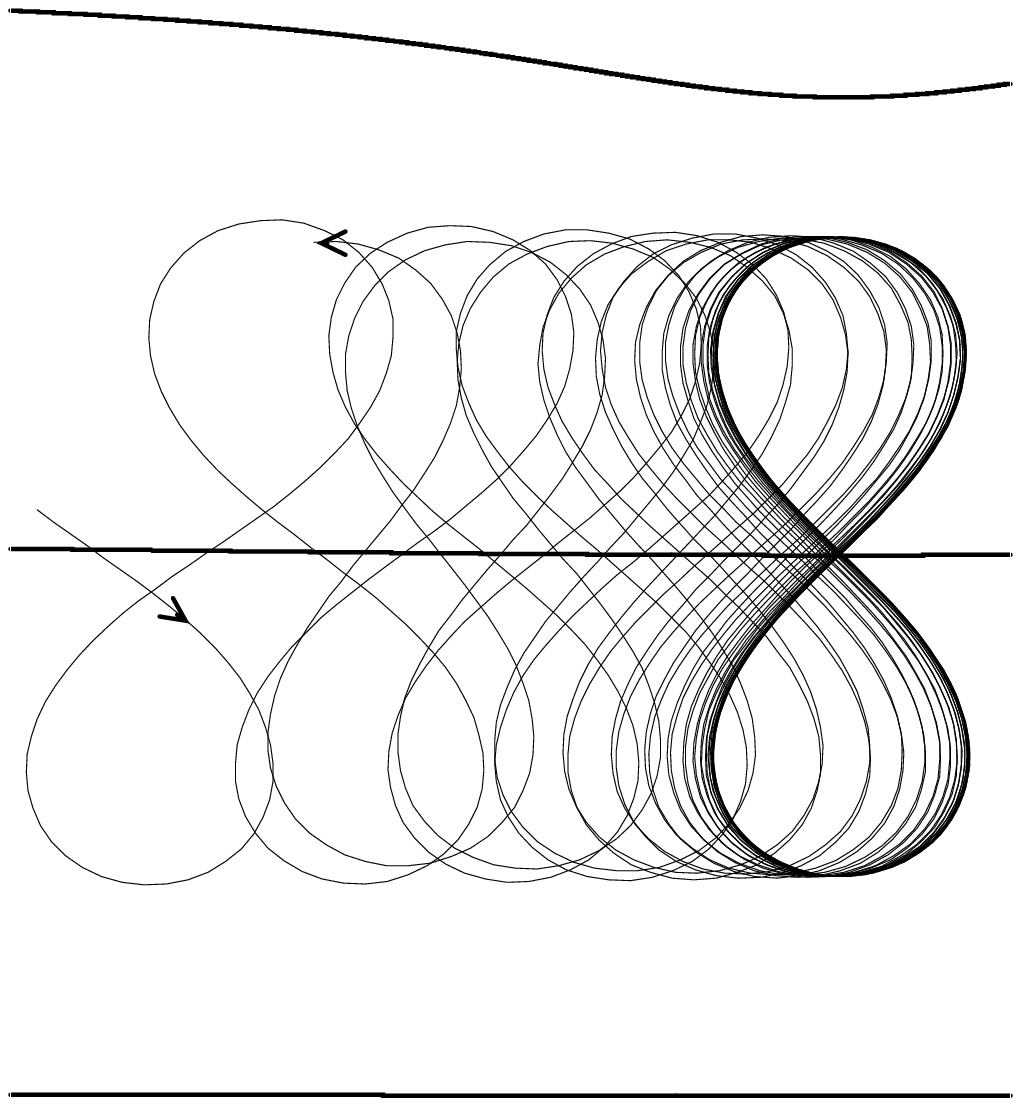}\vspace*{3mm}} 
\caption{``Serpentology''. Top: transformation of a snake state with
large $\theta$ into a drifting orbit with decreasing gradient of the
magnetic field; bottom: reflection of a snake state by a magnetic
bottle-neck.}
\label{fig2}
\end{figure}

\begin{figure}
\centerline{\epsfxsize=72mm\epsfbox{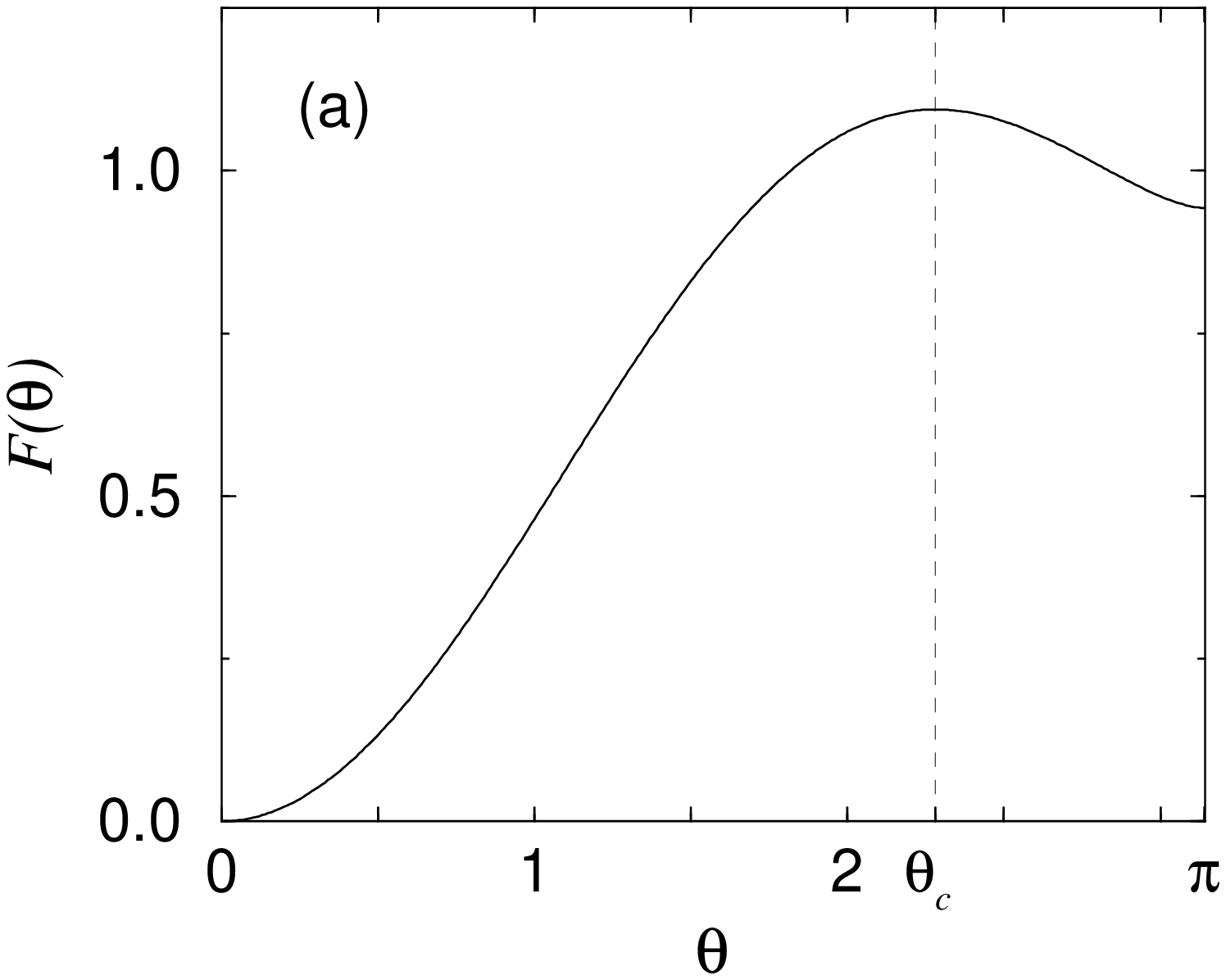}\vspace*{3mm}} 
\centerline{\epsfxsize=72mm\epsfbox{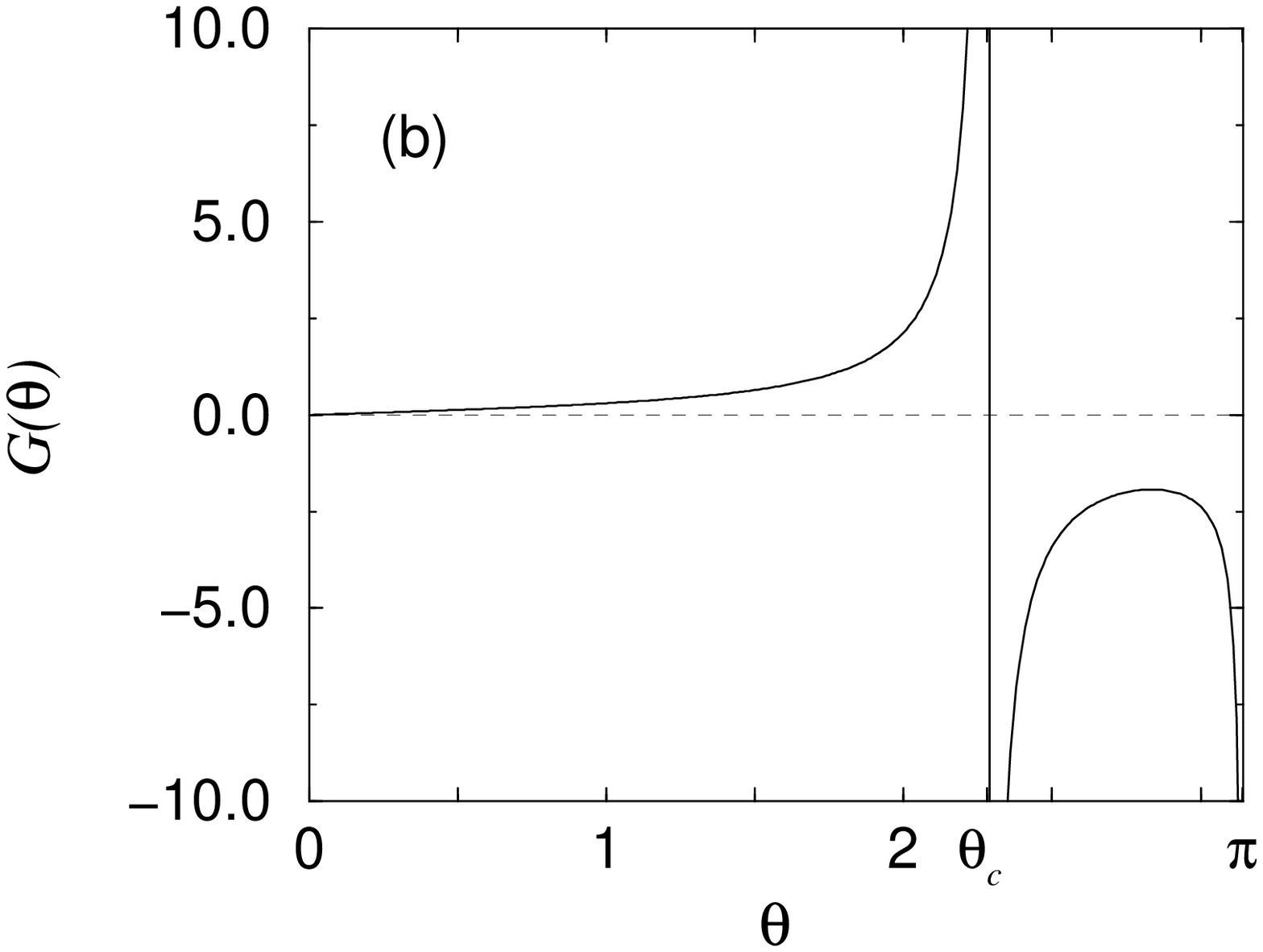}\vspace*{3mm}} 
\centerline{\epsfxsize=72mm\epsfbox{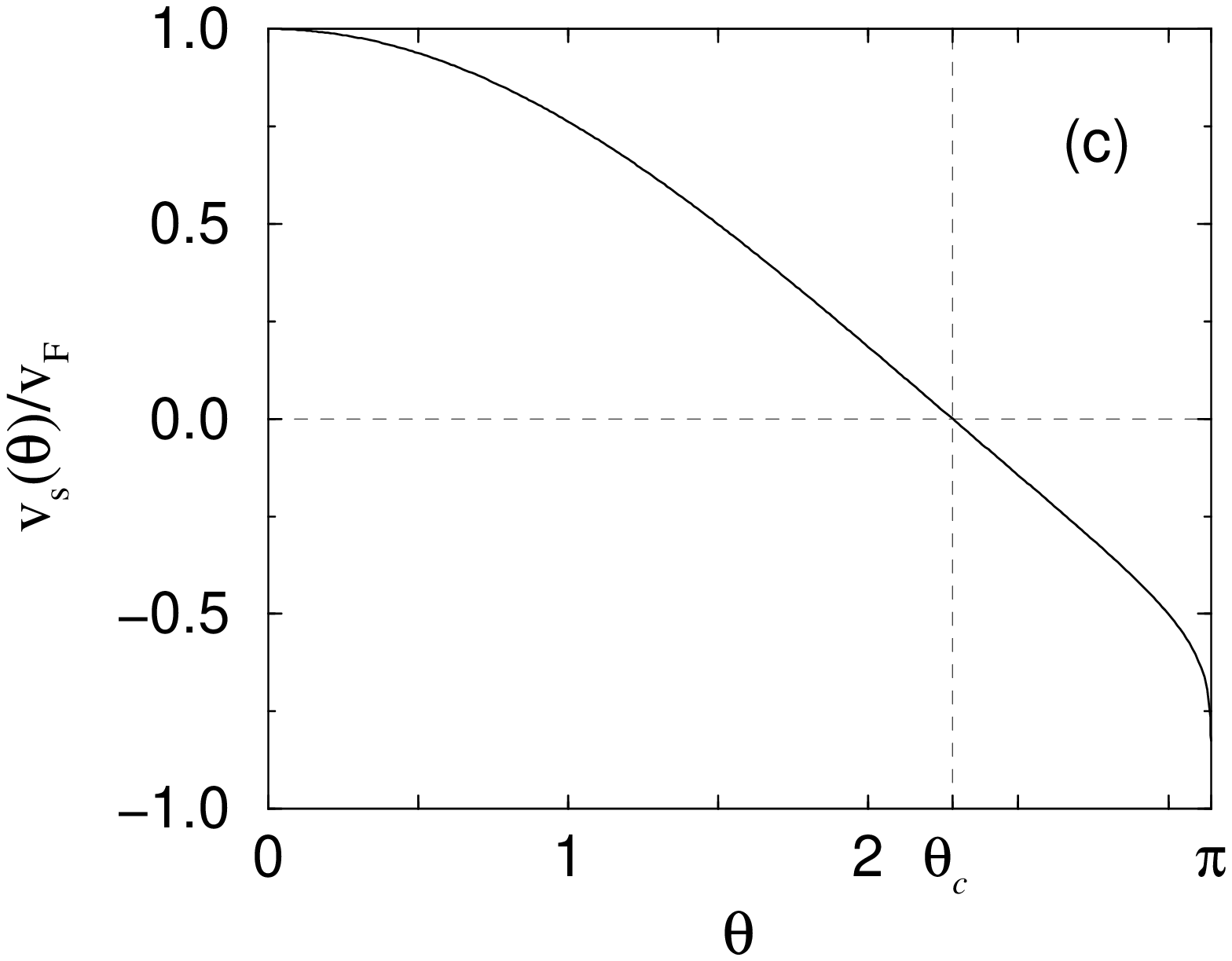}\vspace*{3mm}} 
\caption{The functions $F(\theta)$ and $G(\theta)$ determining the
adiabatic dynamics of the snake states according to
Eqs.~(\ref{e3})--(\ref{e5}), and the snake state velocity $v_s(\theta)$.}
\label{fig3} 
\end{figure}

Everywhere except in small regions near the saddle points, the motion
along the rapidly
oscillating snake-state trajectories around the zero-$B$
contours conserves the adiabatic invariant (see also Ref.~\cite{lee94})
\begin{equation}
\label{e2} 
I_\bot=m\oint\dot ydy~.
\end{equation} Here
$y$ and $\dot y$ are the distance and the velocity in the direction
perpendicular to the zero-$B$ line, and the integral is taken over one
period of the oscillations. The conservation of the quantity $I_\bot$
can be established directly by considering the evolution of the angle
$\theta(x)$ the snake-state trajectory forms with the line of zero
field ($y=0$) at position $x$ along this line as a consequence of the
smoothly varying gradient $b(x)=|\partial B(x,y)/\partial
y|_{y=0}$. The 
adiabatic invariant is parametrized as 
\begin{equation}
\label{e3}
I_\bot(b,\theta)=4mv_F^{3/2}(2mc/eb)^{1/2}F(\theta)~,
\end{equation}
where $F(\theta)$ is a dimensionless function of order unity which can
be found explicitly: 
\begin{equation}
\label{e4}
F(\theta)=(1-\cos\theta)\int_0^1\!\!\!d\xi\sqrt{(1-\xi^4)+
\cos\theta(1-\xi^2)^2} ~.  
\end{equation} 
Note that $I_\bot(b,\theta)$
may be written also as $(e/c)\Phi(b,\theta)$, where $\Phi$ is the
magnetic flux through the area encircled by the snake-state trajectory
and the zero-$B$ line in one oscillation period. 
We represent the equation $dI_\bot/dx=0$ in the
form of a scaling relation for the snake-state angle 
\begin{equation}
\label{e5}
{d\theta\over d\ln b}=G(\theta)~,\quad
G^{-1}(\theta)=2{d\over d\theta}\ln F(\theta)~. 
\end{equation} 
This
equation expresses the adiabatic invariance in terms of the fact that,
given boundary conditions [$\theta (x_0)$ and $b(x_0)$ at some point
$x_0$], the angle $\theta$ at a point $x$ of the trajectory is
completely determined by the gradient $b(x)$ at the same point.  Eq.\
(\ref{e4}) gives the asymptotic expressions for $G(\theta)$:
\begin{eqnarray} 
\label{e6} 
G(\theta)&\simeq&{\theta\over 4}~,\quad
\theta\to 0~;\\ 
\label{e7} 
G(\theta)&\simeq& -{2\over 3}{1\over
(\pi-\theta)\ln {1\over \pi-\theta}}~,\quad \theta\to\pi~.
\end{eqnarray} 
Equations (\ref{e5}), (\ref{e6}) tell us that
$\theta(x)$ obeys the scaling 
\begin{equation} 
\label{e8}
{\theta(x_1)\over \theta(x_2)}=\left[{b(x_1)\over b(x_2)}\right]^{1/4}
\end{equation} 
in the limit of small harmonic oscillations $\theta\to
0$. The singularity of $G(\theta)$ in the opposite limit of
$\theta\to\pi$ is a precursor of the bifurcation which accompanies the
break away of the trajectory from the zero-$B$ line at $\theta=\pi$
(see Fig.~\ref{fig2}, top). 
The functions $F(\theta)$ and $G(\theta)$ in the whole
range of $\theta$ are shown in Fig.~\ref{fig3}a,b. 

The remarkable point to notice
is that $G(\theta)$ changes sign at some $\theta=\theta_c$ (which is
$\simeq 131^\circ$). More specifically, $G(\theta)$ behaves singularly
around $\theta_c$, as $(\theta-\theta_c)^{-1}$, which corresponds to a
maximum in $F(\theta)$ at this point. This behavior of
$I_\bot(\theta)$ means that the velocity of the snake states
$v_s(\theta)$ (which is the average of $\dot x$ over one period) must
change sign at $\theta=\theta_c$, i.e., the snake state is
``reflected" at the point $x_c$ defined by the equation
$\theta(x_c)=\theta_c$ (Fig.~\ref{fig2}, bottom). Indeed, as follows from Eq.\
(\ref{e3}), the constancy of $I_\bot[\theta(x)]$ cannot be maintained
on {\it both} sides of the point $x_c$. Note also that, in terms of
the time evolution of $\theta$, the change in sign of the function
$G(\theta)$ at $\theta=\theta_c$ means that the time derivative
$\dot\theta$ retains the sign it had before the reflection. In fact,
one can show, by solving the problem with constant gradient $b$
exactly, that 
\begin{equation} 
\label{e9}
v_s(\theta)=v_FF'(\theta){1+{1\over 2}\cos^2\theta\over {3\over
2}\sin\theta F(\theta)+\cos\theta F'(\theta)}~, 
\end{equation} 
i.e., $v_s(\theta)$ interpolates between $v_s(0)=v_F$ and $v_s(\pi)=-v_F$
and vanishes at $\theta=\theta_c$ (see Fig.~\ref{fig3}c). 
It is worth noting that the period
of the oscillations $T_s(\theta)$ increases monotonically with growing
$\theta$: 
\begin{equation} 
\label{e10} 
T_s(\theta)={1\over
mv_F^2}\left({3\over 2}+\cot\theta {\partial
\over\partial\theta}\right)I_\bot(b,\theta)~,
\end{equation} 
i.e.,
$T_s(\theta)$ is equal to $2\pi(mc/ebv_F)^{1/2}$ at $\theta=0$ and
diverges as $4(mc/ebv_F)^{1/2}\ln[1/(\pi-\theta)]$ at
$\theta\to\pi$. The ``wavelength" of the snake states along the
direction of propagation obviously reads $\Delta x=|v_s|T_s$, while
the amplitude of the oscillations in the perpendicular direction is
given by $\Delta y=2v_F(mc/ebv_F)^{1/2}\sin(\theta/2)$.

\subsection{Snake-state percolation}
\label{ss2c}

\begin{figure}
\centerline{\epsfxsize=80mm\epsfbox{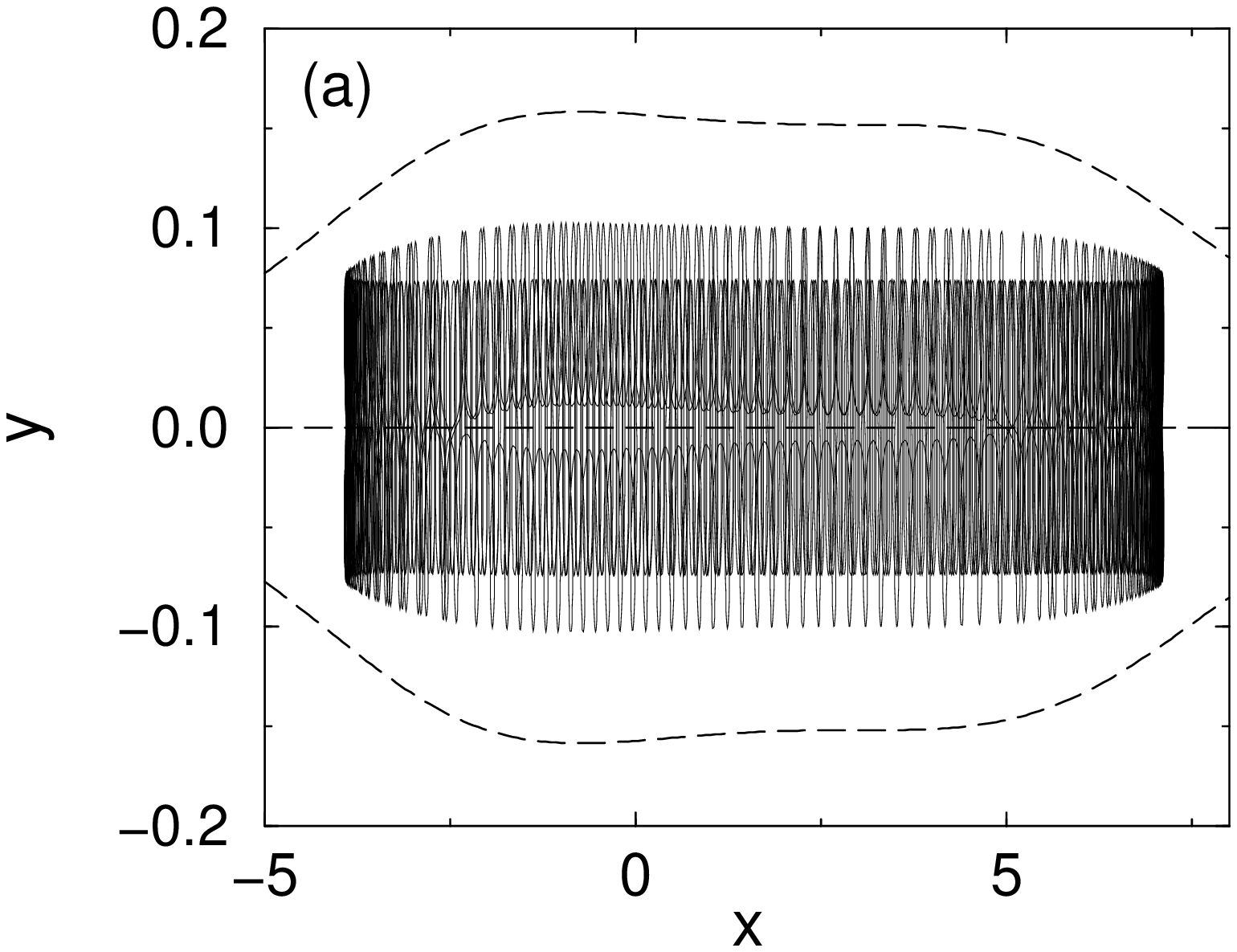}} 
\centerline{\epsfxsize=80mm\epsfbox{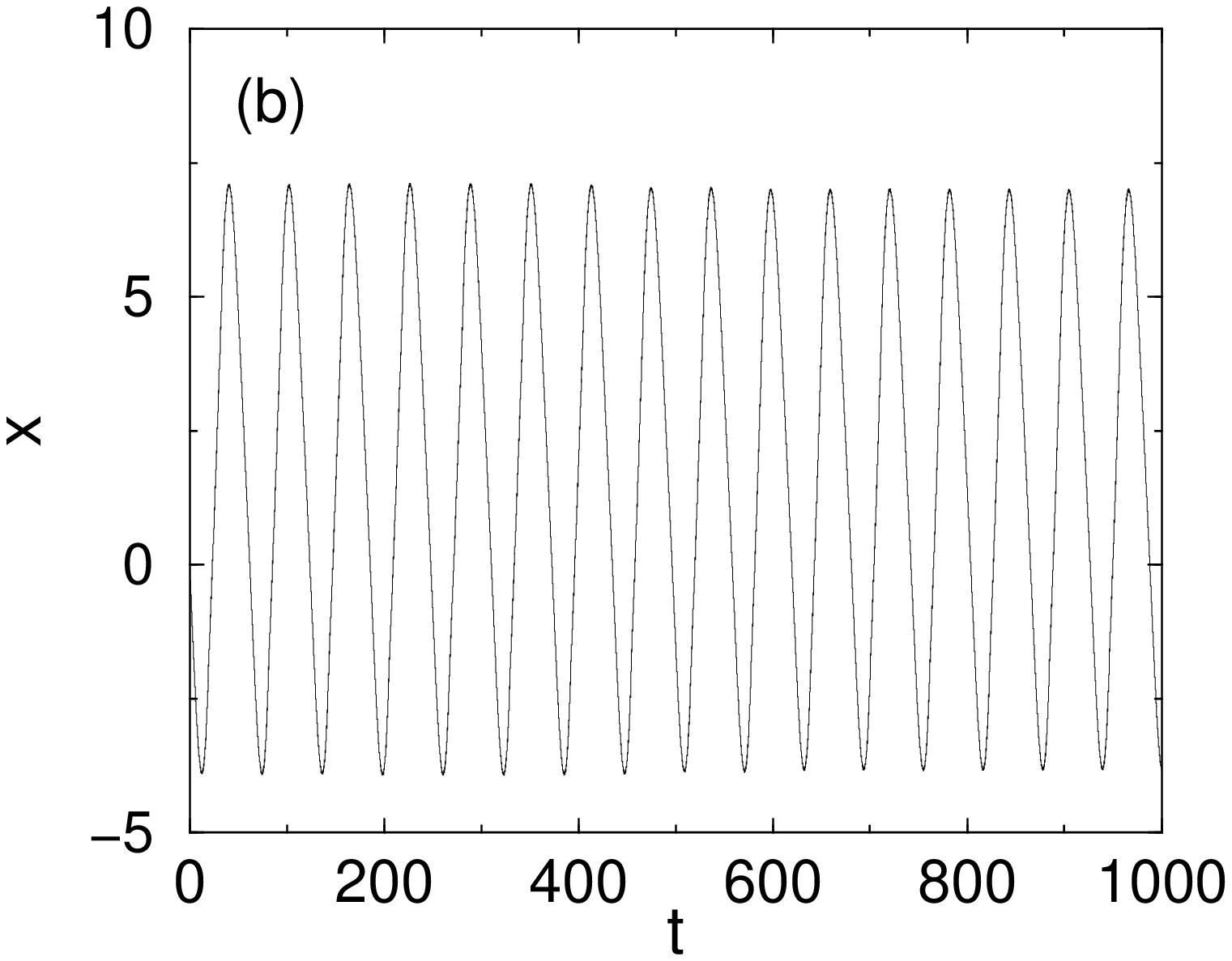}} 
\vspace{0.5cm}
\caption{Snake state in a trap. (a) Real-space trajectory of
a particle trapped between two bottle-necks. The scales
of the $x$ and $y$ axes differ by a factor $\approx 25$:
the figure is ``compressed'' in the $x$-direction. The dashed lines show
the contours of the constant magnetic field. (b)
Time evolution of the $x$ coordinate. It is seen that the drift motion
in the trap is almost periodic.}
\label{fig4} 
\end{figure}

The adiabatic nature of the snake-state dynamics means that a typical
trajectory is ``trapped'' between two return points $x_+$ and $x_-$
with $\theta(x_\pm)=\theta_c$ (Fig.~\ref{fig4}a). Within the adiabatic
picture,  the drift motion  in such a trap is periodic in time, as
demonstrated in Fig.~\ref{fig4}b.  Hence, unless
nonadiabatic corrections are taken into account, these trajectories do
not contribute to the dc conductivity. The nonadiabatic corrections
for a typical trajectory with a slowly varying $\theta(x)$ are
exponentially weak, so that the motion remains finite on an
exponentially long time scale. Yet, this does {\it not} mean that
$\sigma_{xx}(\alpha)$ is exponentially suppressed in the limit of
large $\alpha$. The point is that there are rare (but not
exponentially rare) places along the zero-$B$ contours where the
adiabatic picture fails {\it completely}. These are regions where the
contours pass near the saddle points of $B({\bf r})$.

\begin{figure}
\vspace{-0.2cm}
\centerline{\epsfxsize=80mm\epsfbox{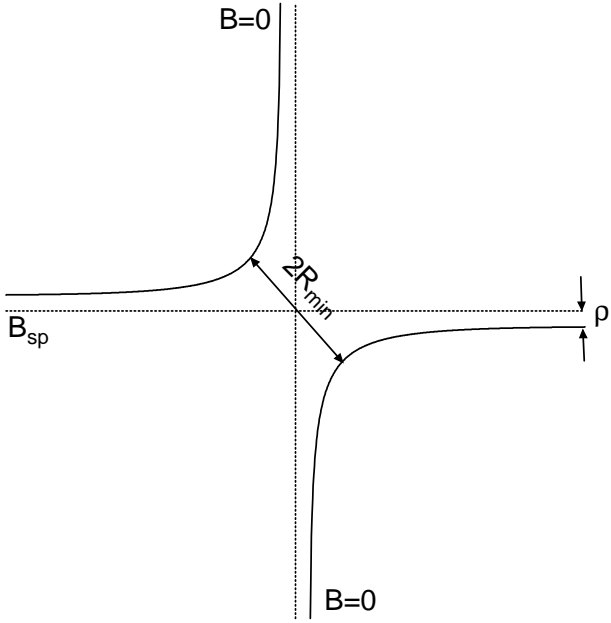}} 
\vspace{0.5cm}
\caption{Geometry of a saddle-point.}
\label{fig5} 
\end{figure}

Consider a snake state that is incident on a saddle point with the
impact parameter $\rho$ (Fig.\ 5). This means that the magnetic field
at the saddle point is $B_{\rm sp}\sim B_0\rho/d$ and the distance
$R_{\rm min}$ at which the zero-$B$ contour passes the saddle point is
$R_{\rm min}\sim \sqrt{d\rho}$. At the saddle point, there is an
intersection of two lines of constant $B({\bf r})=B_{\rm sp}$, while
two zero-$B$ lines, along which the snake states can propagate, come
within the distance $2R_{\rm min}$ from each other. Clearly, if
$R_{\rm min}$ is small enough, the snake state can change the
zero-field contour. The angle $\theta$, which characterizes the type
of the snake-state trajectory, is then also changed, i.e., the
adiabaticity will be broken down upon ``scattering" on the saddle
point. To understand the parameters, consider first the case of
$\rho=0$ (``direct hit"). The snake state propagates then along a
straight line with decreasing gradient $b(x)\sim B_0x/d$, where $x$ is
measured from the saddle point. According to Eq.\ (\ref{e8}),
$\theta(x)\propto x^{1/4}$ decreases when the particle approaches the
saddle point, while, as follows from Eq.\ (\ref{e10}), the wavelength
$\Delta x$ diverges as $x^{-1/2}$. The adiabatic picture is valid only
as long as $\Delta x(x)$ is much smaller than the scale on which the
magnetic field changes, i.e., $\Delta x(x)\ll x$ near the saddle
point, which gives the condition $x\gg x_c$, where $x_c$ is the
characteristic wavelength of the ``last" oscillation before the
particle hits the saddle point. For typical trajectories with
$\theta\sim 1$ and $\Delta x\sim (dR_0)^{1/2}$ at $x\sim d$, this
condition fails at $x\alt x_c\sim d/\alpha^{1/3}$. Now, as is clear
from Fig.\ 6, whether the particle will be scattered to the left of
the saddle point or to the right is determined by the initial phase of
the snake-state oscillations. This sensitivity to the phase signals
the breakdown of the adiabaticity.

\begin{figure}
%\centerline{\epsfxsize=72mm\epsfbox{m1200.1k4.ps}\vspace*{3mm}} 
%\centerline{\epsfxsize=72mm\epsfbox{m1500.1k4.ps}\vspace*{3mm}} 
\centerline{\epsfxsize=72mm\epsfbox{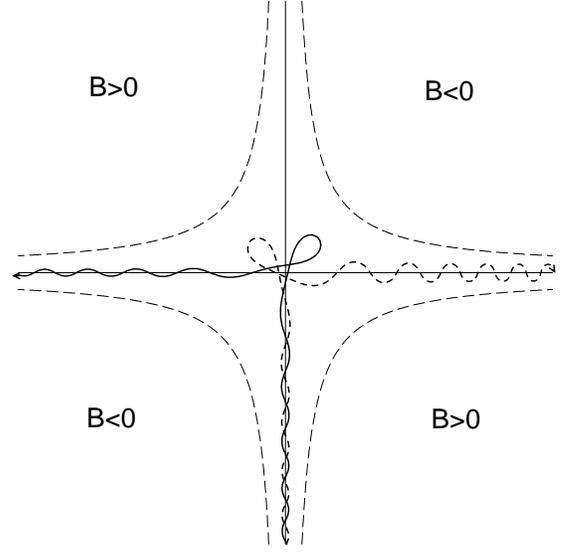}\vspace*{3mm}} 
\caption{Scattering of a snake state at a saddle point. The
particle may turn either left or right, depending on the initial
conditions.} 
\label{fig6} 
\end{figure}

We now turn to the case of finite $\rho$. At large enough $\rho$, the
typical snake-state trajectory does not change the zero-$B$ line: the
condition is that the angle $\theta(x_c)$, with which the trajectory
comes to the saddle point, be much smaller than the ratio $R_{\rm
min}/x_c$. Substituting $\theta(x_c)\sim (x_c/d)^{1/4}$ [see Eq.\
(\ref{e8})], we get $\rho\gg d/\alpha^{5/6}$. However, the adiabatic
invariance is broken at the saddle point in a  wider range of
$\rho$: the condition for the curvature of the zero-$B$ line to be
large on the scale of the wavelength is $R_{\rm min}\alt x_c$, which
gives $\rho\alt\rho_s$, where 
\begin{equation}
\label{e11}
\rho_s\sim
d/\alpha^{2/3}~.
\end{equation} 
Within this range, the angle $\theta$ after the scattering,
$\theta_{\rm out}$, is typically of order unity even though the
particle is incident on the saddle point with a small $\theta$
(moreover, $\theta_{\rm out}$ depends strongly on the phase of the
oscillations of the incoming trajectory). 

Now consider how the particles propagate between such non-adiabatic saddle
points. The saddle points with $\rho\alt\rho_s$ are distributed
sparsely along the zero-$B$ trajectories with the linear density $\sim
\rho_s/d^2$. Therefore, only a small fraction of the snake states can
escape the adiabatic traps on their way between two such saddle
points: most trajectories are localized in between. The snake state is
not trapped if $\theta(x)<\theta_c$ everywhere on its trajectory
between the collisions with the saddle points. According to Eq.\
(\ref{e8}), this is possible for trajectories with sufficiently small
$\theta$. Indeed, consider a snake state which has a small angle
$\theta\ll 1$ in a typical place with the gradient $b\sim
B_0/d$. Typically, it will be able to travel a long distance, by far
larger than $d$, until its angle reaches the value $\theta_c$: this
will occur in a fluctuation of the magnetic field with the anomalously
high gradient $b\sim B_0/d\theta^4$. Since the probability $p(b)$ that
the gradient at a given point exceeds some value $b$ is determined by
the Gaussian statistics, $p(b)=\exp(-b^2/\left<b^2\right>)$, we find
that the state will typically run ballistically the distance
$L(\theta)$ obeying the equation $L(\theta)p(B_0/d\theta^4)/d\sim 1$,
which gives $L(\theta)\sim d\exp(\theta^{-8})$. Hence, using
$L(\theta_s)/d \sim d/\rho_s \sim \alpha^{2/3}$, we conclude that
the states with
the angles $\theta\ll\theta_s$, where
\begin{equation}
\label{e12}
\theta_s\sim
(\ln\alpha)^{-1/8}~,
\end{equation} 
will typically get through to the saddle point.

We thus conclude that the particles with $\theta\alt\theta_s$
propagate between saddle points ``ballistically" with the longitudinal
velocity $v_s\simeq v_F$, while others are simply out of play. Now we
turn to construct the overall picture of the snake-state
propagation. The scattering on a saddle point is actually a multi-step
process. The fact that the angle $\theta_{\rm out}$ is typically $\sim
1$ means that, having collided with a saddle point once, the particle
almost inevitably returns back to it with a new angle of incidence
$\theta'$: in effect, the trajectory ``sticks" to the saddle
point. However, after many mappings $\theta\to\theta_{\rm
out}\to\theta'$, the multiple reflections establish a stochastic
distribution of the angle $\theta_{\rm out}$ characterizing the
outgoing trajectory. Also, after many attempts the particle will go to
the left or to the right with equal probability. This randomization of
$\theta_{\rm out}$ and of the direction of motion is clearly seen in
the numerical simulation in \cite{lee94}. Once the particle picks up
the angle $\theta_{\rm out}\sim \theta_s/\alpha^{1/12}$, it will move
ballistically until it reaches 
the next saddle point. Here, the factor
$\alpha^{-1/12}$ is related to the fact that the angle $\theta(x)$
will increase $\propto x^{1/4}$ on the scale of $d$, so that the
particle must have $\theta_{\rm out}$ which is $(x_c/d)^{1/4}$ times
smaller than $\theta_s$. At the new saddle point the whole process
will repeat itself. Since the saddle points are separated by the large
distance $\sim d^2/\rho_s$, the average
time it takes the particle to move to
the next saddle point is determined by the ballistic propagation between
them, which requires the time $t_b\sim d^2/\rho_sv_F\sim
(d/v_F)\alpha^{2/3}$, 
 not by the multiple attempts to ``break away" with large
$\theta_{\rm out}$, which end in  returns to the starting
point. Indeed, assuming the full randomization of $\theta_{\rm out}$,
we estimate the number of such attempts, until the particle picks up
the angle $\theta_{\rm out}\lesssim\theta_{\rm
out}^{(c)}=\alpha^{-1/12}(\ln \alpha)^{-1/8}$ necessary to reach the next
saddle-point, as $N\sim 1/\theta_{\rm out}^{(c)}$. According to
what is said above, the initial condition $\theta_{\rm out}$ 
allows the particle to advance the distance
\begin{equation}
\label{e12a}
L(\theta_{\rm out})\sim d \times \left\{ \begin{array}{ll}
\alpha^{-1/3}\theta_{\rm out}^{-4}\ , & \ \ \ \theta_{\rm
out}\gtrsim\alpha^{-1/12}\\ 
\exp[(\alpha^{1/12}\theta_{\rm out})^{-8}]\ , & \ \ \ 
\theta_{\rm out}\lesssim\alpha^{-1/12}\ .
\end{array} \right.
\end{equation}
Therefore, the average ``waiting time'' the particle spends in the
unsuccessful attempts to reach the next saddle point is
\begin{equation}
\label{e12b}
t_w\sim N\int_{\theta_{\rm out}^{(c)}} d\theta_{\rm out} L(\theta_{\rm
out})/v_F \sim (d/v_F)\alpha^{2/3}/\ln\alpha\ .
\end{equation}
Thus, the total time it takes to get through from one saddle point to
another is indeed determined by $t_b\gg t_w$.

\subsection{Conductivity in a strong random magnetic field}
\label{ss2d}

Now let us calculate the conductivity at $\alpha\gg 1$. As was
mentioned at the very beginning, most trajectories do not contribute
to $\sigma_{xx}$ since they follow periodic drift orbits. Next, we
turned to consider a special class of the trajectories -- the snake
states. However, as we showed above, most snake states are also
localized in the adiabatic traps and only those with angles smaller
than $\theta_s$ can propagate along the lines of zero $B$. At this
point, we have to be concerned about the topology of the zero-$B$
contours. The first thing to notice is that all the contours are
closed except one and this one percolating contour by itself cannot
yield a finite conductivity. The conductivity is nonetheless finite
since the snake states in fact form a conducting {\it network} of
finite width. The nodes of the network are {\it critical} saddle
points, where two adjacent percolating contours come close to each
other. Note that most of the saddle points that the particle hits on
its way between the critical ones only connect up small closed loops
and so do not create a connected network. This happens only at the
critical saddle points, where the snake states can cross over from one
critical zero-$B$ line of length $L_s\sim \alpha^{14/9}d$ to
another. We use here the results of the percolation theory (for a
review see, e.g., \cite{isichenko92}): $L_s\sim d(d/\rho_s)^{\nu+1}$,
where $\nu=4/3$ is the critical exponent that controls the size of the
critical cluster $\xi_s\sim d(d/\rho_s)^\nu$, so that the ratio of the
length of the trajectory and the distance from the starting point
$L_s/\xi_s\sim d/\rho_s$. The characteristic distance between the
nodes, i.e., the size of the elementary cell $\xi_s$, is then
$d\alpha^{8/9}$. On length scales longer than $\xi_s$, the particle
dynamics can be viewed as fully stochastic. We estimate the
macroscopic diffusion coefficient as $D\sim\nu_sD_s$, where $\nu_s\sim
L_sR_s\theta_s^2/\xi_s^2$ is the fraction of particles residing in the
delocalized snake-states and $D_s\sim\xi_s^2\times v_F/L_s$ is their
diffusion coefficient. Note that $\nu_s$ contains a factor
$\theta_s^2$ --
since the density of the snake states is determined in the phase space
parametrized by both the angle $\theta$ and real-space coordinate:
accordingly, one factor $\theta_s$ comes from the calculation of the
fraction of the plane covered by the conducting snake states, while
the other describes their fraction in the $\theta$ space. We thus have
$D\sim v_FR_s\theta_s^2$ and, correspondingly \cite{ln},
\begin{equation} 
\label{e13} 
\sigma_{xx}\sim {e^2\over
h}{k_Fd\over\alpha^{1/2}{\cal L}}~,\quad{\cal L}\sim
\ln^{1/4}\alpha,\quad\alpha\gtrsim 1~.  
\end{equation} 
It is worth noting that the percolation {\it enhances} the conductivity:
by comparison with the Born approximation [Eq.\ (\ref{e1})], the
conductivity is $\sim\alpha^{3/2}/{\cal L}$ times larger (though the
localization effects are strong and naively one might have expected
the opposite). Let us also note that $\sigma_{xx}$ given by Eq.\
(\ref{e13}) is larger by a factor of $\sim\alpha^{1/2}/{\cal L}$ than
that obtained for $\alpha\gg 1$ in \cite{khveshchenko96} by using an
``eikonal approach''. The fault in \cite{khveshchenko96} is not with
the quasiclassical approximation itself, but with the method of
disorder averaging, which neglects the localization of particles and
the percolating character of the transport through the snake states.

\begin{figure}
\centerline{\epsfxsize=72mm\epsfbox{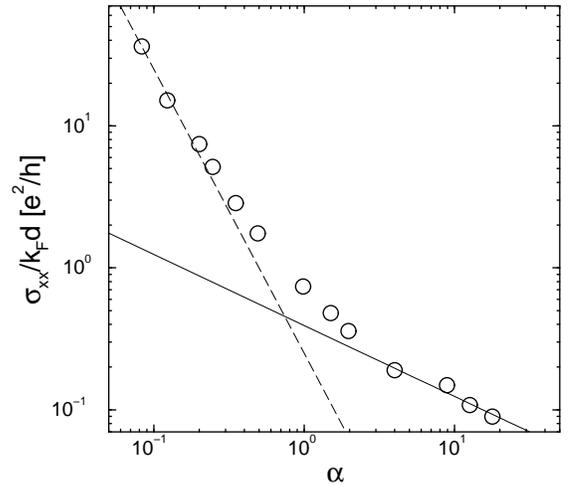}\vspace*{3mm}} 
\caption{{\it dc} conductivity at $\overline B=0$, as a function of the
parameter $\alpha$. The dashed and the full lines correspond to the
theoretical asymptotics (\ref{e1}) and (\ref{e13}), respectively.
Statistical errors do not exceed the symbol size.} 
\label{fig7} 
\end{figure}

We now turn to the numerical simulation. 
To calculate the conductivity tensor components $\sigma_{\mu\nu}$ we
evaluate numerically the classical current response function,
\begin{equation}
\label{e13a}
\sigma_{\mu\nu}=e^2\rho_F\int_0^\infty dt\,\langle v_\mu(0)
v_\nu(t)\rangle \ ,
\end{equation}
where $\rho_F=m/2\pi\hbar^2$ is the density of states and
the average is taken over the disorder realizations and starting
points of the trajectory. Typically, evaluation of the conductivities
involved averaging over $\sim 10^3\div 10^4$ trajectories. 
The numerical results for
$\sigma_{xx}$ in Fig.~\ref{fig7} fully confirm the analytical findings
above. For small $\alpha$, the results are in good agreement with the
Born approximation formula, Eq.\ (\ref{e1}), while at $\alpha\sim 1$ a
crossover to the $\alpha^{-1/2}$ behavior, Eq.\ (\ref{e13}), takes
place. At $\alpha=1/\sqrt{2}$ (the value relevant to the CF problem at
$n=n_i$ and in the absence of impurity correlations) we find
$\sigma_{xx}\simeq 1.0 (e^2/h)k_F d$, which is a factor of $\sim 2$
larger than the Born approximation value. This improves the agreement
with the experimentally found CF conductivity (defined as the inverse
of the measured resistivity at $\nu=1/2$), though the typical
experimental values of $\sigma_{xx}$ are still larger than the one we
obtain by a factor of $\sim 2-3$. This remaining discrepancy might be
attributed to correlations in the distribution of the charged donors
\cite{correl}, which reduces the effective strength of the random
potential and thus reduces $\alpha$. The resistivity data in zero
external magnetic field (as contrasted to zero effective magnetic
field acting on CFs) indeed indicate that the model of statistically
independent impurity positions overestimates the amount of disorder
\cite{coleridge91,correl}. It is also worth noting here, in
view of the controversy about the effective mass of the CFs
\cite{halperin93,effmass,du94,coleridge95}, that in the RMF model
$\sigma_{xx}$ at zero $\overline 
B$ [Eqs.\ (\ref{e1}),(\ref{e13})] does not depend on $m$ (neglecting
the corrections \cite{interact}
related to the interaction between the CFs).

\section{dc transport in non-zero mean magnetic field}
\label{s3}

We now consider the conductivity at finite $\overline B$. Let us first
discuss the case of $\alpha\sim 1$, when the conductivity $\sigma
({\overline B})$ can be parameterized as a function of the {\it
single} variable ${\overline B}/B_0$. As shown in the previous
section, at small ${\overline B}/B_0$ we are at the crossover between
the uncorrelated diffusion and the snake-state percolation. Now, at
${\overline B}\gg B_0$ the particle dynamics is a slow van Alfv\'en
drift of the cyclotron orbits along the lines of constant $\delta B
({\bf r})$. It follows that the conductivity is determined by a
percolation network of trajectories close to the $\delta B ({\bf
r})=0$ lines.  From the point of view of topology of the network, the
situation is thus similar to that at zero $\overline B$ and $\alpha\gg
1$. What is crucially different, however, is the mechanism of the
percolation. Specifically, at large $\overline B$ there is no
stochastic mixing at the nodes of the percolation network: unlike the
snake states at ${\overline B}=0$, the rapidly rotating cyclotron
orbits pass unharmed through the critical saddle points of $\delta
B({\bf r})$ without crossing over to the adjacent cell.  In the
high-$\overline B$ limit, the mixing occurs on the links of the
network and is only due to the weak scattering between the drift
trajectories.

In order to calculate the conductivity at ${\overline B}\gg B_0$, we
should integrate out the fast cyclotron rotation, taking care not to
lose the effect of the nonadiabatic mixing. Specifically, we have to
go beyond the standard separation of the fast and slow degrees of
freedom, known as the drift approximation.  The parameter that governs
this separation is $\delta/d$, where $\delta$ is a characteristic
shift of the guiding center after one cyclotron revolution. The drift
approximation is represented as a power series in $\delta/d\ll 1$. In
our problem at $\alpha\sim 1$, this parameter is the ratio
$B_0/{\overline B}$. Therefore, if ${\overline B}\gg B_0$, the
adiabatic description is good on {\it microscopic} scales. The key
point, however, is that the conductivity is strictly zero at the level
of the drift approximation -- since the drift orbits are periodic in
the thermodynamic limit. The effects that break the adiabatic
invariance and lead to the transitions between the drift orbits are
{\it exponentially} weak at $\delta/d\ll 1$.

\subsection{Single-impurity scattering}
\label{ss3a}

The problem of the scattering between the drift trajectories in the
static RMF \cite{mirlin98}, as well as a similar problem for a random
scalar potential, considered recently in \cite{fogler97}, is a
particular example of the broad class of problems dealing with
non-conservation of an adiabatic invariant. Despite the general
interest of this problem, any systematic
expansion capable of giving the scattering rate {\it beyond} the
exponential accuracy, has proven to be a tough exercise. To consider a
transparent example, we formulated and solved parametrically exactly a
{\it single}-scattering problem \cite{evers98}. Specifically, we
introduce a weak homogeneous gradient $\epsilon$ of the background
magnetic field and consider the interaction with an ``impurity"
modeled by a spatially localized perturbation $\delta B({\bf r})$ of
size $d$, so that the total field 
\begin{equation} 
\label{e14} 
B({\bf r})={\overline B}[1+\epsilon (y/R_c)]+\delta B({\bf r})~,
\end{equation} 
where $R_c$ is the cyclotron radius in the field
$\overline B$. The guiding center coordinate $y$ averaged over the
cyclotron orbit, $\rho =\left< y\right>_c$, plays the role of an
impact parameter. The particle entering the system at $x=-\infty$ with
$\left< y \right>_c=\rho_i$ will leave it at $x=\infty$ along the
trajectory with $\left< y\right>_c=\rho_i+\Delta\rho$, where
$\Delta\rho$ is the nonadiabatic shift we are interested in. In this
single-impurity scattering problem, the shift is a perfectly well-defined
quantity. To first order in $\delta B$, the exact solution is given by
\begin{equation} 
\label{e15} 
\Delta\rho=\gamma I~,\quad
I=\int_{-\infty}^{\infty}dt {\delta B[{\bf r}_0(t)]\over {\overline
B}}\dot y_0(t)~.  
\end{equation} 
Here ${\bf r}_0(t)$ is the
unperturbed trajectory for $\delta B=0$, 
\begin{equation}
\label{e16}
\gamma(\epsilon)=\omega_c\left<\dot y_0(t)\int_0^t dt'\dot
x_0(t')/\dot y_0^2(t')\right>_c~,
\end{equation} 
$\omega_c=e{\overline B}/mc$, 
and the angular brackets denote averaging over one cyclotron
period. In the limit $\epsilon\to 0$ 
the constant $\gamma\to -1$. We can further
simplify the model by assuming that $\delta B$ is a function of $x$
only. In this case, the integral in Eq.\ (\ref{e15}) is evaluated at
$\epsilon\ll 1$ by the saddle-point method to give 
\begin{eqnarray}
\label{e17} 
\Delta\rho&=&-{2v_F\over {\overline
B}}\sqrt{\epsilon\over\pi}\cos\left({2\over\epsilon}- {\pi\over
4}\right)\cos \varphi_0\\ \nonumber &\times&\int_{-\infty}^\infty dt
\cos \omega_ct\,\,\delta B\left({\epsilon\over 2}v_Ft-R_c\right)~.
\end{eqnarray} 
This formula expresses the nonadiabatic shift in terms
of the asymptotics of the Fourier transform of the smooth function
$\delta B(x)$ -- thus demonstrating explicitly the exponential
smallness of $\Delta\rho$. It shows that the parameter which governs
the exponential falloff of $\Delta\rho$ is $d/\delta\gg 1$, 
where $\delta=\pi\epsilon R_c$,  while the
ratio $d/R_c$ may be arbitrary. Note that the pre-exponential factor
happens to oscillate wildly as $\epsilon\to 0$. These oscillations are
geometric resonances due to the commensurability of two length scales
$R_c$ and $\delta$. Remarkably, the series of the resonances is
defined by the properties of the unperturbed solution
(``self-commensurability") and not by the shape of the scatterer. This
means that the oscillations are damped with increasing strength of the
perturbation \cite{evers98} -- since the resonance condition cannot be
met simultaneously everywhere on a strongly perturbed
trajectory. Another peculiar feature of the nonadiabatic shift is its
sensitivity to the phase $\varphi_0$ of the cyclotron rotation of the
incident electron ($\Delta\rho\propto\cos\varphi_0$).

In the CF problem, a charged impurity located at a distance $d$ from
the plane occupied by the electron gas creates the axially symmetric
perturbation $\delta B({\bf r})=\delta B_0d^3(r^2+d^2)^{-3/2}$.
Because of the branch points at $r=\pm id$ in this expression, the
integrand in Eq.\ (\ref{e15}) will contain the exponentially small
factor $\exp \left(-{2\omega_c\over\epsilon
v_F}\sqrt{[y_0(t)-\rho_i]^2+d^2}\right)$. The lengthy general result
reduces to \begin{eqnarray} \label{e18} \Delta\rho &\simeq& 8\pi
d\,\,{\delta B_0\over {\overline B}}\,\,{\sqrt{R_cd}\over \delta} \\
\nonumber &\times&\cos\varphi_0\cos\left({2\over\epsilon}-{\pi\over
4}\right)\exp\left(-{2\pi d\over\delta}\right) \end{eqnarray} at
$\rho_i=0$ in the limit $d\gg R_c$. Equation (\ref{e18}) reflects the
features of the non-adiabatic shift discussed above: the exponential
smallness, the oscillations with changing $\epsilon$, and the
oscillatory dependence on the phase $\varphi_0$. These results were
confirmed by numerical simulations in Ref.~\cite{evers98}. Note that
Eq.\ (\ref{e18}) implies that the drift trajectory is only slightly
perturbed by $\delta B({\bf r})$.
 
\subsection{Optimum fluctuation}
\label{ss3b}

In the transport problem one has to average over an ensemble of
impurities. What is crucial for the averaging process is that the
drift velocity is itself determined by the fluctuations of the
impurity field. The non-linear problem gets therefore much more
involved as compared to the single-scattering model above, but the
principal features of the nonadiabatic scattering remain unchanged and
the main message can be simply stated: Because of the exponentially
strong dependence of the shift on the parameters of the single
scatterer, the conductivity is determined by {\it rare} places with an
anomalously high rate of nonadiabatic transitions
\cite{fogler97,mirlin98}.  Accordingly, one can neglect correlations
between consecutive transitions from one drift orbit to another and
each nonadiabatic shift can be considered independently. Since the
nonadiabatic scattering rate increases as the drift motion gets
faster, the effective scatterers, sparsely distributed along the
percolating trajectories, are characterized by anomalously sharp
changes of the RMF. The problem now is to find the density and the
parameters of these scatterers.

The nonadiabatic shift $\Delta\rho=\Delta\rho_x+i\Delta\rho_y$ (in
complex notation) after one scattering reads
\begin{equation}
\label{e18a} 
\Delta\rho=v_F\int dt
e^{i\omega_ct+i\varphi_0}A(t)~, 
\end{equation} 
where the smooth
function $A(t)$ varies slowly on the scale of $\omega_c^{-1}$ and is
given by the following average taken over one cyclotron period:
\begin{equation} 
\label{e18b} 
A(t)=\left<e^{i\chi(t)}\right>_c~,\quad
\chi(t)={e\over mc}\int_0^tdt'\delta B[{\bf r}(t')]~. 
\end{equation}
The integral which determines the random phase $\chi(t)$ should be
done on the exact trajectory ${\bf r}(t)$. Note that Eq.\ (\ref{e18a})
gives the nonadiabatic shift both along and across the drift
trajectory. Since only the latter is of interest, one should project
the result of the integration (\ref{e18a}) onto the axis perpendicular
to the direction of the drift of the outgoing particle.

Since the nonadiabatic mixing is determined by the short wavelength
Fourier harmonics of the perturbation [Eq.\ (\ref{e18a})], it is the
analytical properties of the function $A(t)$ and, therefore, of the
correlator $\left<\delta B(0)\delta B({\bf r})\right>$ that are
important. In the CF problem, this correlator has branch points as a
function of $r$ at $r=\pm 2id$. However, in order to calculate the
scattering probability, which is given by Eq.\ (\ref{e18a}), one has
to find the singularities in $\delta B[{\bf r}(t)]$ as a function of
{\it time} $t$ and average the result. For a given perturbation
$\delta B({\bf r})$ this purely mechanical problem of finding the
Fourier asymptotics of the integral along the path ${\bf r}(t)$ may be
quite complex, but we can circumvent the difficulties by performing
the configurational averaging first. As was already mentioned, the
effective scatterers are characterized by anomalously large
fluctuations of the drift velocity 
\begin{equation}
\label{e18+1}
v_d(s)=v_F(R_c/2B)|\nabla B(s)|~,
\end{equation} 
where $s$ is
the coordinate along the path. To see this, one can use the exact
solution of the single-scattering problem considered above. Let us
first assume, for the purpose of illustration, that the large $v_d(s)$
does not change appreciably on the scale of $d$. Equation (\ref{e18})
then tells us that a single impurity located on the trajectory which
passes through the fluctuation with large $v_d$ yields
$\Delta\rho(v_d)\propto \exp (-d\omega_c/v_d)$. One sees that
$\Delta\rho(v_d)$ grows exponentially with increasing $v_d$. Now, the
linear density of the fluctuations with large $v_d$ along the
percolating path is of order $p(v_d)$, where the Gaussian
probability that the drift velocity at a given point is larger than
$v_d$ reads 
\begin{eqnarray}
\label{e18c}
&& p(v_d)=\exp (-v_d^2/2\left<v_{dx}^2\right>); \nonumber \\
&& \left<v_{dx}^2\right>={3\over 16}v_F^2\left({R_c\over
d}\right)^2\left({B_0\over {\overline B}}\right)^2\ .
\end{eqnarray}
Averaging $[\Delta\rho(v_{d})]^2$ with $p(v_d)$, we thus get
$$\left<[\Delta\rho(v_d)]^2\right>\propto\exp (-S_{\rm min})\ ;\qquad
S_{\rm min}=3\left({d^2\omega_c^2\over
2\left<v_{dx}^2\right>}\right)^{1/3}.$$ 
The  ``optimum'' drift velocity which determines this average is
$v_d^0=\left<v_{dx}^2\right>^{1/2}(4d^2\omega_c^2/
\left<v_{dx}^2\right>)^{1/6}$. As is clear, $v_d^0\gg
\left<v_{dx}^2\right>^{1/2}$ at ${\overline B}\gg B_0$. The optimum
fluctuations yield the Gaussian behavior of the scattering rate:
\begin{equation}
\label{e22} 
S_{\rm min}=c({\overline B}/B_0)^2~,\quad {\overline
B}/B_0\gg 1~, 
\end{equation} 
with the coefficient $c=18^{1/3}\simeq 2.62$ (here we assumed
$\alpha=1/\sqrt{2}$). 
This simple derivation of the exponential dependence of
$\left<[\Delta\rho(v_d)]^2\right>$ captures the essential physics and
yields a correct parametric estimate for $S_{\rm min}$; however, it is not
exact in that $v_d(s)$ in the optimum fluctuation is not, in fact,
constant on the scale of $d$, and for this reason it does not give the
correct value of the numerical coefficient $c$ in (\ref{e22}). 
To obtain the asymptotically exact
numerical coefficient in $S_{\rm min}$, we have to use the
optimum-fluctuation method in the whole configurational space. The
optimum configuration is characterized by the function $v_d(s)$ and
the shape of the drift trajectory. We write the phase factor
$e^{i\omega_c t}$ in \mbox{Eq.\ (\ref{e18a})} as $e^{i\varphi (s)}$,
where we introduce the $s$ dependent phase 
\begin{equation}
\label{e18+2} 
\varphi(s)=\omega_c\int_0^s{ds'\over
v_d(s')}~,
\end{equation} 
and notice that the exponent $S_{\rm min}$ is
determined by the phase $\varphi (s)$ picked up at the singular point
of the perturbation $\delta B[{\bf r}(t)]$ regarded as a function of
the longitudinal coordinate $s$. As can be verified by varying the
shape of the trajectory, the minimum ``action" $S_{\rm min}$ is
acquired along a straight path and the quantity to be calculated is
therefore 
\begin{equation} 
\label{e19} 
S_{\rm min}=-\ln \left<\exp
\left(i\omega_c\int_{-id}^{id}{ds\over v_d(s)} \right)\right>~,
\end{equation} 
where the integral should be done along the straight
line connecting the points $s=-id$ and $s=id$ in the complex plane of
the variable $s$ \cite{id}. This average determines, with 
exponential accuracy, the diffusion coefficient $D_\bot\propto\exp
(-S_{\rm min})$, which is defined as 
\begin{equation} 
\label{e20}
D_\bot=\lim_{t\to\infty}t^{-1}\left<\left[\Delta\rho
(t)\right]^2\right>~, 
\end{equation} 
where $\Delta\rho(t)$ is the
nonadiabatic shift across the percolating drift trajectory in time
$t$. The exponent can be written as a sum of two terms, $S_{\rm
min}=W_1+W_2$, where 
\begin{eqnarray} 
\label{e21} 
W_1 &=& {1\over 2}\int\!\!\!{d^2{\bf q}\over (2\pi)^2}
v^0_{dx_{\bf q}} \Gamma_{\bf q}^{-1} v^0_{dx_{-{\bf q}}} \nonumber \\
W_2 &=& i\omega_c\int_{-id}^{id}\!{dx\over
v_{dx}^0(x,0)}\ ,
\end{eqnarray}
and
\begin{equation}
\label{e21b}
\Gamma_{\bf q}=\left<v_{dx}v_{dx}\right>_{\bf
q}=\left({h c^2 mv_F^2\over e^2 {\overline
B}^2}\right)^2nq_y^2e^{-2qd} 
\end{equation}
is the Fourier transform of the drift-velocity correlation function
$\Gamma({\bf r})=\left<v_{dx}(0)v_{dx}({\bf r})\right>$.
Here $W_1$ determines the probability
for the optimum fluctuation $v^0_{dx}({\bf r})$ to occur, while $W_2$
describes the nonadiabatic scattering on this fluctuation. 
The variational equation $\delta W/\delta v^0_{dx}=0$ yields
\begin{equation} 
\label{e21a} 
v_{dx}^0({\bf r})= i\omega_c\int_{-id}^{id}{dx'\over
\left[v_{dx}^0(x',0)\right]^2}\Gamma(x-x',y)~,
\end{equation} 
which is a non-linear integral equation for $v_{dx}^0(x,0)$ with
$x\in(-id,id)$. Its solution defines, by means 
of analytical continuation, the
optimum fluctuation on the real axis of $x$. Dimensional analysis of
this equation shows that the solution has the form $v^0_{dx}({\bf
r})=v_F(B_0/\overline{B}){\cal G}({\bf r}/d)$, where $\cal G$ is a
dimensionless function of order unity, which leads again to
Eq.~(\ref{e22}).  
To find the exact value of $c$, one has to determine the function
${\cal G}$, which requires solving the integral equation  (\ref{e21a}). 
We used a variational approach,
choosing the trial function $v_{dx}^0({\bf r})=\kappa \Gamma({\bf r})$ 
with the variational parameter $\kappa$. This trial function is the
optimal fluctuation for the (slightly different) 
problem of finding a large value
$v_{dx}^0(0)$ at the point ${\bf r}=0$
and should give a good estimate for $c$. The result is
$c\simeq 2.28$, close to the value found above within the
simplified consideration neglecting the spatial variation of $v_{dx}$ on
the scale of $d$.

\subsection{Magnetotransport at $\alpha\sim 1$:
Conductivity of the composite fermions at $\nu\neq 1/2$}
\label{ss3c}

We are now prepared to calculate the conductivity at ${\overline B}\gg
B_0$. 
Similarly to the case of the snake-state percolation, the nonadiabatic
transitions create a conducting network with the elementary cell of
size $\xi\sim d(d/R_d)^{4/3}$ and perimeter $L\sim \xi d/R_d$, where
$R_d$, the width of the links of the network composed of the drift
trajectories, obeys the equation \begin{equation} \label{e23}
R_d^2\sim D_\bot L/v_d~.\end{equation} This equation is the condition
of connectivity of the network. The conductivity due to the
nonadiabatic mixing of the drift trajectories is thus given by
\begin{equation} 
\label{e24} 
\sigma_{xx}\sim {me^2\over
\hbar^2}v_dR_d\propto {e^2\over h}k_F d
\,\exp \left(-{3\over 13}S_{\rm min}\right)~.
\end{equation} 
The dissipative transport is seen to be
strongly suppressed beyond the scale $\overline B\sim B_0$. Let us
emphasize that the crossover to the adiabatic regime occurs in the CF
system at a small deviation from half-filling $\nu=1/2$: ${\overline
B}\sim B_0$ corresponds to the shift $|\nu -1/2|\sim 1/k_Fd\ll 1$.

\begin{figure}
\centerline{\epsfxsize=75mm\epsfbox{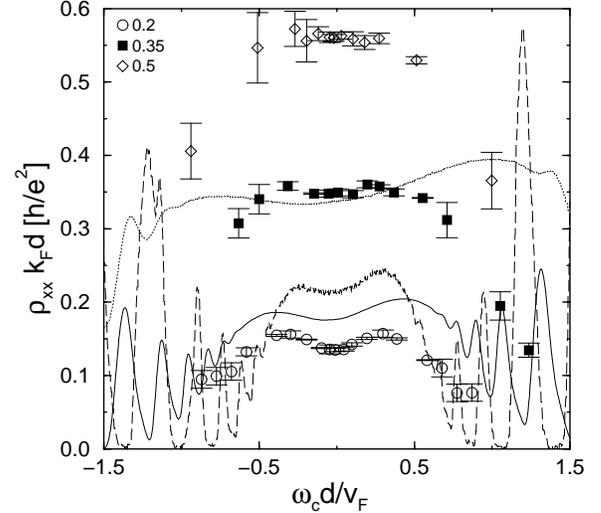}} 
\caption{Magnetoresistivity at $\alpha= 0.2$, 0.35, and 0.5; 
the cyclotron frequency in the scaling of the $x$-axis refers to the
external magnetic field, $\omega_c=e{\overline B}/mc$. The full, 
dashed, and dotted 
lines show the experimental magnetoresistivity around
$\nu=1/2$  according to the data of Refs.~\protect\cite{smet_unpubl}, 
\protect\cite{stormer}, and
\protect\cite{coleridge95}, respectively (in this case $\omega_c$
refers to the effective magnetic field $\overline B=B-2hcn/e$). The
sample parameters (carrier density $n$, undoped spacer width $d$ and
zero-field mobility $\mu$) are:  
$n=1.53\times 10^{11}\:{\rm cm}^{-2}$, $d=80\:{\rm nm}$, $\mu=5\times
10^6\:{\rm cm}^2/V\,s$ \protect\cite{smet_unpubl},
$n=0.86\times 10^{11}\:{\rm cm}^{-2}$, $d=80\:{\rm nm}$, $\mu=10.5\times
10^6\:{\rm cm}^2/V\,s$ \protect\cite{stormer},
$n=1.31\times 10^{11}\:{\rm cm}^{-2}$, $d=120\:{\rm nm}$, $\mu=3.5\times
10^6\:{\rm cm}^2/V\,s$ \protect\cite{coleridge95}.}
\label{fig8} 
\end{figure}

Calculation of the conductivity tensor components at ${\overline
B}\lesssim B_0$ cannot be done analytically at $\alpha\sim 1$. 
We have performed extensive numerical simulations to study the overall
shape of the magnetoresistivity. The results for $\alpha=0.2$, 0.35,
and 0.5 are shown in Fig.~\ref{fig8}, in comparison with experimental
data on the magnetoresistivity in the vicinity of $\nu=1/2$ from
Refs.~\cite{coleridge95,stormer}. We recall, that although a simple model
of uncorrelated impurities with the concentration $n_i=n$ gives
$\alpha=1/\sqrt{2}$, the actual value of $\alpha$ may be somewhat
smaller because of impurity correlations. As is seen, the experimental
curves are reasonably close to the numerical data with $\alpha\sim
0.25\div 0.35$. The numerical results show a  positive
magnetoresistance at ${\overline B}\lesssim B_0$, which is followed at
larger ${\overline B}$ by a fall-off in agreement with the analytical
prediction (\ref{e22}). Both these features are clearly observed in
the experiment. Better agreement between the theory and the experiment
is hardly possible, taking into account the
difference between the experimental
data obtained by different groups and on different (though nominally
very close) samples. Sufficiently far from
half-filling the experimental curves start to show 
magnetooscillations, which have not been included in the 
classical model above. 
We will discuss the issue of the magnetooscillations below.

\subsection{Magnetotransport at weak disorder, $\alpha\ll 1$.}
\label{ss3d}

We now turn to consider the case of weak disorder $\alpha\ll 1$. The
Gaussian behavior of the conductivity at large $\overline B$ is in
fact a general property of the nonadiabatic transport in the RMF. We
can use the same optimum fluctuation method as for the CF system above
to get 
\begin{equation} 
\label{e25} 
S_{\rm min}= c(2\alpha^2)^{2/3}\left({{\overline
B}/B_0}\right)^2~,\quad {\overline B}/B_0\gg \alpha^{-1}~. 
\end{equation} 
This result is valid if
$\overline B$ is large enough; namely, it requires that the shift of
the guiding center in the field of the optimum fluctuation after one
cyclotron revolution, $\delta$, and the Larmor radius $R_c$ be both
smaller than $d$. The condition $R_c\ll d$ breaks down with decreasing
$\overline B$ at ${\overline B}/B_0\sim \alpha^{-1}$. Provided
$\alpha\ll 1$, the exponent (\ref{e25}) taken at the crossover point
is large, which means that $\delta\ll d$ and we are still deep in the
adiabatic regime. Next we consider the regime ${\overline
B}/B_0<\alpha^{-1}$. 
What changes at smaller $\overline B$ is that the
drift velocity of the guiding center at point ${\bf r}$ is now
determined by an effective RMF $\delta B^{\rm eff}({\bf r})$, the
amplitude of which is smaller than $B_0$ because of the averaging of
the fluctuations over the large cyclotron radius. We define $\delta
B^{\rm eff}({\bf r})$ by writing the general expression for the drift
velocity to first order in $\delta B/{\overline B}$ in the form
\begin{eqnarray} 
\label{e26} 
{\bf v}_d({\bf r})&=&-{e\over
mc}\int_0^{2\pi}{d\varphi\over 2\pi}\left[\delta {\bf B}({\bf r}+{\bf
R}_{\varphi})\times {\bf R}_\varphi\right]\\ \nonumber &=&
v_F{R_c\over 2{\overline B}^2}\left[\nabla\delta B^{\rm eff}({\bf
r})\times {\overline {\bf B}}\right]~, 
\end{eqnarray} 
where ${\bf
R}_\varphi=R_c(\cos\varphi,\sin\varphi)$ and $\varphi$ is the phase of
the cyclotron rotation around the point ${\bf r}$. Stokes' theorem
then yields 
\begin{equation} 
\label{e27} 
\delta B^{\rm eff}({\bf
r)}={1\over \pi R_c^2}\int_{R\leq R_c}d^2{\bf R}\delta B({\bf r}+{\bf
R})~.  
\end{equation} 
Equation (\ref{e27}) tells us that $\delta B^{\rm eff}$
is given by the magnetic flux through the cyclotron orbit, so that the
drift occurs along the lines of constant {\it flux} (not the lines of
constant magnetic field averaged along the orbit). If $\delta B ({\bf
r})$ is a smoothly varying function on the scale of $R_c$, $\delta
B^{\rm eff}({\bf r})$ coincides with $\delta B ({\bf r})$, otherwise
the averaging leads to a strong suppression of the fluctuations of
$\delta B^{\rm eff}({\bf r})$. More specifically, at $R_c\gg d$ the
field $\delta B^{\rm eff}({\bf r})$ is characterized by two spatial
scales: it has a short-range component whose correlation radius
remains of order $d$ -- its characteristic amplitude is
\begin{equation}
\label{e28} 
B^{\rm eff}_0\sim
B_0(d/R_c)^{3/2}
\end{equation} 
-- and a long-range component which has
a larger amplitude $\sim B_0d/R_c$ but fluctuates on the much longer
scale of $R_c$. Since the drift velocity is given by the gradient of
$\delta B^{\rm eff}({\bf r})$, it is the short-range fluctuations that
determine ${\bf v}_d({\bf r})$. Applying the optimum fluctuation
method, we now have 
\begin{equation} 
\label{e29} 
S_{\rm min}\sim
\alpha^{1/3} {\overline B}/B_0~,\quad \alpha^{-1/3}\alt {\overline
B}/B_0\alt \alpha^{-1}~.  
\end{equation} 
At ${\overline B}/B_0\sim
\alpha^{-1/3}$ the factor $e^{-S_{\rm min}}$ becomes of order unity
and at smaller $\overline B$ the transport is described, in the first
approximation, in terms of 
the conventional Drude theory; namely, the conductivity grows with
decreasing $\overline B$ as $\sigma_{xx}\sim k_Fd(B_0/{\overline
B})^2$ until it saturates at ${\overline B}/B_0\sim\alpha$ at the
level given by Eq.\ (\ref{e1}). 

Let us briefly discuss the (weak) deviations from the Drude behavior
at low magnetic fields ${\overline B}$. The Drude theory predicts zero
magnetoresistance, $\Delta\rho_{xx}/\rho_{xx}=0$. The low-field
magnetoresistance at $\alpha\ll 1$ was studied in
\cite{mirlin98,khveshchenko96}, where it was found that 
\begin{equation}
\label{e29a} 
{\Delta\rho_{xx}\over\rho_{xx}}=-3
\alpha^2|\ln\alpha|{{\overline B}^2\over B_0^2}\ ,\qquad {\overline
B}\ll\alpha B_0\ .
\end{equation}
This result is valid in the weak magnetic field regime
$\omega_c\tau_{tr}\ll 1$ (which is equivalent to the condition 
${\overline B}\ll\alpha B_0$). As our numerical results indicate, this
(very weak) negative magnetoresistance crosses over to a much more
pronounced positive magnetoresistance in the intermediate range 
$\alpha B_0\ll{\overline B}\ll\alpha^{-1/3} B_0$, so that the
resistivity shows a maximum at ${\overline B}\sim\alpha^{-1/3} B_0$
before it starts to drop exponentially according to Eqs.~(\ref{e24}),
(\ref{e25}), (\ref{e29}). This positive magnetoresistance remains for
all $\alpha$ below $\sim 0.5$, see Figs.~\ref{fig8},~\ref{fig9}, and is
strikingly similar to the experimentally observed positive
magnetoresistance of composite fermions near $\nu=1/2$. 
We are currently working on a theoretical explanation of this effect. 

\subsection{Magnetotransport at strong disorder, $\alpha\gg 1$: 
Localization of the snake
states at ${\overline B}>{\overline B}_c$} 
\label{ss3e}

\begin{figure}
\centerline{\epsfxsize=75mm\epsfbox{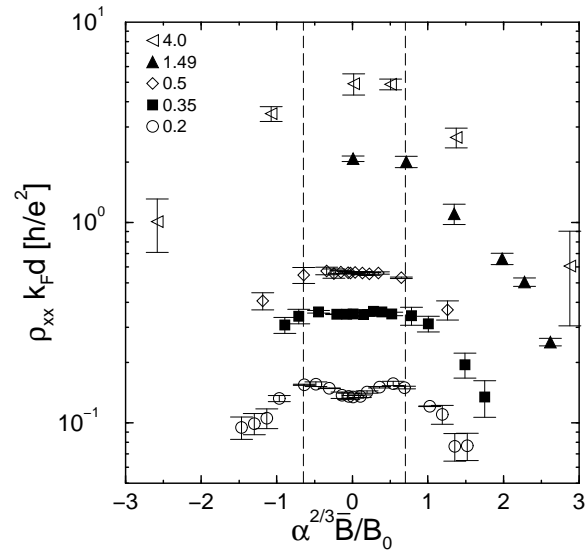}} 
\caption{Magnetoresistivity at $\alpha= 0.2,$ 0.35, 0.5, 1.5, and
4. The dotted lines show the critical field $B_c$, at which the
exponential fall-off begins for $\alpha\protect\gtrsim 1$; see
Eq.~(\ref{e30}). } 
\label{fig9} 
\end{figure}

Let us address now the finite-$\overline B$ transport in the opposite
limit of strong disorder $\alpha\gg 1$. The asymptotics (\ref{e25})
remains unmodified at sufficiently large $\overline B$, namely, at
${\overline B}\gg B_0/\alpha^{1/2}$, where $R_c\ll R_s$. At smaller
$\overline B$, the trajectories close to the zero-$\delta B$ lines
become localized in the adiabatic traps [Sec.\ \ref{s2}] and start to
transform into the snake states. A new feature is thus the appearance
of a competing mechanism of the conduction -- the snake-state
percolation. Indeed, the snake-state trajectories are tied to the
lines of $B({\bf r})=0$, where $B({\bf r})={\overline B}+\delta B({\bf
r})$ is the {\it total} field. On the other hand, the percolating
trajectories are those that follow the lines of $\delta B({\bf r})=0$,
independent of $\overline B$. At large ${\overline B}\gg
B_0/\alpha^{1/2}$, the percolating and snake-state trajectories are
therefore separated in space: the snake states are closed orbits
localized deep inside the elementary cells of the conducting
network. Now, at ${\overline B}\alt B_0/\alpha^{1/2}$, the
nonadiabatic scattering is due to two mechanisms: the exponentially
weak nonadiabatic corrections to the dynamics of the snake-state angle
$\theta$ and the breakdown of the adiabaticity of the snake states at
the saddle points. The latter mechanism leads to the formation of
percolation clusters. Thus, there exists a well-defined value of
${\overline B}={\overline B}_c$ below which the snake states form a
continuous network and can percolate through the entire system. The
critical field ${\overline B}_c$ is of the order of $B_0\rho_s/d$,
i.e., 
\begin{equation} 
\label{e30} 
{\overline B}_c\sim B_0\alpha^{-2/3}~, 
\end{equation} 
which is the characteristic
amplitude of the magnetic field at the critical saddle points. It is
worth stressing that there is a clear separation of the adiabatic and
snake-state regimes: the nonadiabatic scattering within the finite
clusters only leads to an exponentially narrow uncertainty in the
position of the critical point. If one neglects this nonadiabatic
smearing of the transition, the conductivity can be expressed as
\begin{equation} 
\label{e31} 
\sigma_{xx}={e^2\over h}\,{k_Fd\over
\alpha^{1/2}{\cal L}}\,\,{\cal F}\!\left({{\overline B}\over
{\overline B}_c}\right)~, 
\end{equation} 
where ${\cal F}(x)$ is a
dimensionless function, such that ${\cal F}(0)\sim 1$ and ${\cal
F}(x\geq 1)=0$, and $\sigma_{xx}$ at zero $\overline B$ coincides with
that in Eq.\ (\ref{e13}). The magnetoresistivity for $\alpha=1.5$ and
$\alpha=4$ is shown on a logarithmic scale in Fig.~\ref{fig9} (for
completeness, we also included the data for smaller values of $\alpha$,
which have been already displayed in Fig.~\ref{fig8}). We see that the
resistivity indeed shows an exponential fall-off beyond a
characteristic field consistent with Eq.~(\ref{e30}).

To find the critical behavior of $\sigma_{xx}\propto ({\overline
B}_c-{\overline B})^t$ near ${\overline B}_c$, we
formulate an auxiliary percolation problem in more conventional
terms. Consider equipotential contours in a random potential with a
characteristic amplitude $V_0$ and the correlation length $d$ and pick
up all contours within the energy band $(-\Delta,\Delta)$, where
$\Delta\ll V_0$. These contours form a percolation network, the size
of the elementary cell of which is $\xi(\Delta)\sim
d(V_0/\Delta)^{4/3}$. The characteristic width of the links of the
network is of order $d\Delta/V_0$. This is a standard percolation
problem. Now, let us shift the energy band corresponding to the
percolation network: namely, we introduce a parameter $\varepsilon$
and consider all contours within the band
$(-\Delta+\varepsilon,\Delta+\varepsilon)$. Clearly, the system
undergoes a percolation transition at $\varepsilon=\pm\Delta$. In this
new percolation problem, the length $\xi(\Delta)$ plays a role of the
elementary scale, so that the characteristic radius of the percolation
cluster is now
$\xi(\Delta,\varepsilon)\sim\xi(\Delta,0)(\Delta/|\Delta-
\varepsilon|)^{4/3}$ (for positive $\varepsilon$). Next, let particles
propagate along the equipotential contours ``ballistically".
As we know already from Secs.\ \ref{s2},\ref{s3}, the
conductivity $\sigma$ of such a network scales as the typical width of
the conducting links $d(\Delta-\varepsilon)/V_0$, i.e., $\sigma\propto
(\Delta-\varepsilon)$. Assuming that our original problem can be
mapped onto the problem above, we will get the critical exponent for
the conductivity of the snake-state network $t=1$, i.e., ${\cal
F}(x)\sim 1-x$ at $x\to 1$.

\subsection{Magnetooscillations}
\label{ss3g}

Until now we have dealt with the transport properties of {\it
classical} particles. The classical description is justified by the
large value of the parameter $k_Fd\gg 1$: first, this parameter
enabled us to calculate the conductivity by examining the microscopic
dynamics in terms of classical {\it trajectories}; second, it
guaranteed the existence of a wide range of the fields $B_0$ and
$\overline B$ in which the thus calculated conductivity
$\sigma_{xx}\gg e^2/h$ [cf.\ Eqs.\ (\ref{e1}),(\ref{e13})]. It is the
latter condition that allowed us to neglect the quantum interference
of multiply scattered waves and related localization effects.
In this section, we discuss the small quantum
oscillations of $\sigma_{xx}$ as a function of $\overline B$, which are
conventionally termed the Shubnikov-de Haas (SdH) effect. We argue
that the physical picture of the magnetooscillations in the 
smooth random magnetic field (and, in fact, in the limit of a long
ranged random potential as well) is rather peculiar from the conventional
point of view.

We first recall the standard results for the case of a random scalar
potential with a sufficiently short correlation length.
Within the usual approach \cite{ando74,isihara93}, the
magnetooscillations of $\sigma_{xx}$  are related to those of
the total density of states $\rho_F$: $\sigma_{xx}^{\rm
osc}/\sigma_{xx}\propto \rho_F^{\rm osc}/\rho_F$, where $\sigma_{xx}^{\rm
osc}$ and $\rho_F^{\rm osc}$ denote the oscillating parts of
$\sigma_{xx}$ and $\rho_F$. The damping of the oscillations in low
fields is described in terms of the single-particle relaxation time
$\tau_s$ (also termed sometimes the quantum relaxation time)
\cite{shoenberg84,coleridge89,raikh93,laikhtman94,aronov95}, which is
equal to the transport mean free time $\tau_{tr}$ for the white-noise
random potential and is $\sim \tau_{tr}/(k_F d)^2$ for the random
potential with  correlation length $d\gtrsim k_F^{-1}$. Specifically, the
exponent of the Dingle factor $e^{-S}$ is $S=\pi/\omega_c\tau_s$, so
that at
$\omega_c\tau_s\sim 1$ the oscillations of both the density of states
at the Fermi level and the conductivity become strong. We note that
this crossover to the strong oscillations occurs in the case of
short-range disorder at $\sigma_{xx}/( e^2/h)\sim k_F
l(\tau_s/\tau_{tr})^2\gg 1$, assuming that the ratio 
$\tau_{tr}/\tau_{s}$ is not too large. With further
increasing $\overline B$, at $\omega_c\tau_s\gg 1$, the density of
states exhibits a series of peaks, well separated from each other,
so that in the centers of the valleys between the peaks 
the quantum localization starts to develop
rapidly, which leads in turn to the appearance of quantum plateaus in
the Hall conductivity. The full Hall quantization takes place, however,
at much larger ${\overline B}$, when the value of the conductivity in
the center of the peak drops to a value $\sim e^2/h$. Therefore, in
the short-range random potential there exists a parametrically broad
region between the appearance of the SdH oscillations and the fully
developed quantum Hall effect. 

In the case of a  white-noise random magnetic field, the
situation is different. The short-range fluctuations of the magnetic
field are accompanied by long-range fluctuations of the random
vector-potential. As a result,
the single-particle relaxation rate now
diverges due to the strong small-angle scattering
\cite{amw94,aronov95}. The divergence is cut off by the characteristic
length scale in the problem, which is the cyclotron radius.
As a consequence, the damping factor $S$ takes
a different form, $S=4\pi E_F/\omega_c^2\tau_{tr}$ \cite{aronov95}. 
It follows that at $S\sim 1$, where the oscillations
become observable,  $\sigma_{xx}\sim e^2/h$. This
is, therefore, a marginal case: the SdH oscillations become
appreciable in the same region of ${\overline B}$, where the quantum
localization effects get strong.

Now let us consider the limit of the long-range RMF. 
In this case, the oscillations of the total density of states are
damped exponentially strongly in the whole region, where the classical
conductivity is $\gtrsim e^2/h$, and are thus of no importance.
However, the crucial thing
to notice is that the oscillations of $\sigma_{xx}$ and of the total
density of states $\rho_F$ are no longer directly related to each
other. Indeed, as shown above, $\sigma_{xx}$ is determined with
growing $\overline B$ by a progressively smaller fraction of the total
number of trajectories. In the adiabatic limit, only trajectories
close to the zero-$\delta B$ contours contribute to $\sigma_{xx}$. It
follows that the oscillations of $\sigma_{xx}$ are associated with the
oscillations of the density of these conducting states {\it only}. In
particular, this means that the SdH effect and the magnetooscillations
of thermodynamic quantities (de Haas-van Alfv\'en effect) will be
characterized by completely different damping factors.

We now turn to the Dingle factor for $\sigma_{xx}$. Quite generally,
the SdH effect is due to the quantum interference of two waves
propagating along  quasiclassical trajectories for which the number of
cyclotron revolutions is different by $\pm N$ for the $N$th
harmonic of the oscillations
\cite{richter95,hackenbroich95,aronov95}. The Dingle factor for the first 
(and most prominent) harmonic with $N=1$ can therefore be
represented as 
\begin{equation} 
\label{e31a} 
e^{-S}\cos\psi=
\left<\sum_\alpha G_\alpha\cos\Phi_\alpha \right>~, 
\end{equation}
where $\alpha$ labels trajectories in a given realization of disorder,
$G_\alpha$ is the weight with which the trajectory $\alpha$
contributes to $\sigma_{xx}$, $\sum_\alpha G_\alpha=1$, $\Phi_\alpha$
is the phase which is acquired by a particle moving along the
trajectory after one cyclotron revolution, $\psi$ is the phase of the
SdH oscillations, and $\left<\quad\right>$ denotes ensemble averaging.
In general, the phase factor $\cos\Phi$ should be averaged both over
{\it different} trajectories and along {\it one} trajectory. However,
in the limit of smooth disorder, the action $\Phi_\alpha$, which is
the dimensionless magnetic flux through the cyclotron orbit, is the
adiabatic {\it invariant} characterizing the trajectory. A subtle
point here is that the van Alfv\'en drift occurs along the lines of
constant flux $\Phi$ -- not the lines of constant field $B$ [see Eq.\
(\ref{e27})]. At first glance, this difference might seem to be
irrelevant in the case of long-range RMF with $d\gg R_c$. In fact,
however, it is of crucial importance for the calculation of the
amplitude of the oscillations. Let us illustrate this point by first
deriving $S$ along the following line of argument. We know already
that the extended trajectories form a percolating network along the
contours of zero $\delta B({\bf r})$. Since the width of the links of
the conducting network is much smaller (in fact, as shown above,
exponentially smaller) than $R_c$, it is a good approximation to place
the guiding center on the contour of zero $\delta B({\bf r})$,
calculate the RMF flux $\delta \Phi$ through the cyclotron orbit, and
average $e^{i\delta\Phi}$ over different positions of the guiding
center on the zero-$\delta B$ line. This would give a contribution to
the Dingle-factor exponent $S=\left<(\delta \Phi)^2\right>/2\sim
(eB_0/\hbar c)^2(R_c^2/d)^4$ \cite{mirlin98}. In fact, however, the actual
trajectory of the guiding center is slightly shifted from the contour
of zero $\delta B({\bf r})$ -- by an amount which exactly cancels the
above contribution to $S$ -- since it is the flux that is the
adiabatic invariant. It follows that the average over the flux should
be done only over {\it different} trajectories: $\Phi_\alpha$ in Eq.\
(\ref{e31a}) is, in the adiabatic limit, constant for a given
$\alpha$.

Having established the conservation of the flux $\Phi_\alpha$ along
the trajectory we can rewrite Eq.\ (\ref{e31a}) in the following form
\begin{equation} 
\label{e31b} 
e^{-S}\cos\psi=\int d\Phi
G(\Phi)\cos\Phi~, 
\end{equation} 
where $G(\Phi)=\left<\sum_\alpha
G_\alpha\delta (\Phi-\Phi_\alpha )\right>$ is understood as the
ensemble average taken {\it across} a link of the conducting
network. The function $g(\Phi)$ is represented as a narrow peak of
width $\Delta\Phi$ centered at $\Phi_0=2\pi
(mv_F^2/2\hbar\omega_c)$. The dimensionless magnetic flux $\Delta\Phi$
can be expressed in terms of the width $R_d$ [Eq.\ (\ref{e23})] of the
link of the conducting network by $\Delta\Phi\sim (e/\hbar c)\,\Delta
B\, R_c^2$, where $\Delta B\sim B_0R_d/d$ is the characteristic
change of the RMF across the link. We thus see that the broadening
$\Delta\Phi$ is related to the conductivity of the network [Eq.\
(\ref{e24})] and obeys the equation 
\begin{equation} 
\label{e31c}
\sigma_{xx}\sim {e^2\over h}\Delta\Phi~.  
\end{equation} 
This rather
remarkable result implies that the Dingle factor $e^{-S}$ is a function of
the single variable $g=\sigma_{xx}/(e^2/h)$: 
\begin{equation}
\label{e31d}
e^{-S(g)}=\int_{-\infty}^\infty dx Q(x)\cos (2\pi gx)~. 
\end{equation}
Here, we have expressed $G(\Phi)=(2\pi g)^{-1}Q[(\Phi-\Phi_0)/2\pi g]$
in terms of the parameterless function $Q(x)$ which falls off at
$|x|\sim 1$ and is normalized according to $\int dxQ(x)=1$. According
to Eq.\ (\ref{e31d}), the Dingle factor is represented as 
the Fourier transform of the smooth function $Q(x)$. It
is worth noting  that Eq.\ (\ref{e31d}) can be interpreted also in
terms of the local Landau levels. In the language of the
quasiclassical quantization, the contribution $G_N$ of the $N$th Landau
level to the conductivity $\sigma_{xx}=\sum_NG_N$ falls off beyond the
band of width $\Delta N\sim g\gg 1$ around $N=
mv^2_F/2\hbar\omega_c$, where the number of effectively conducting
Landau levels is determined by the change of the flux across the link,
$\Delta N=\Delta\Phi/2\pi$. With this fact taken into account,
Eq.~(\ref{e31c}) takes the familiar form: the conductivity is of the
order of $e^2/h$ times the number of the conducting channels in the
effective network. Applying Poisson's formula to the sum
$\sum_NG_N$ we again arrive at Eq.\ (\ref{e31d}).

\begin{figure}
\centerline{\epsfxsize=75mm\epsfbox{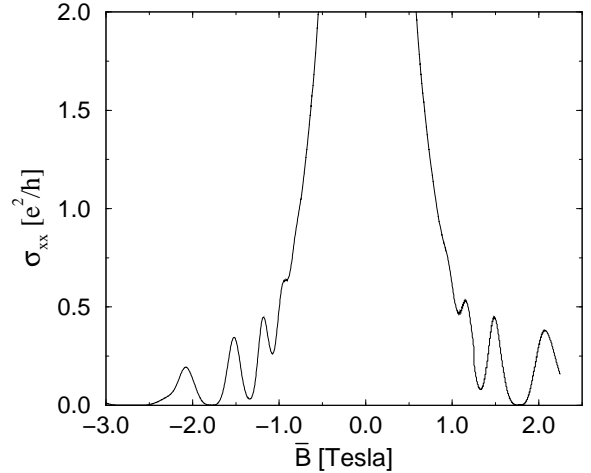}} 
\caption{Composite fermion conductivity $\sigma_{xx}$ as a function of
the effective magnetic field extracted from the data of
 Ref.~\protect\cite{coleridge95}. The magnetooscillations start to
develop when the conductivity (in units of $e^2/h$) drops down to a
value $\sim 1$.}  
\label{fig9a} 
\end{figure}

Hence, the SdH oscillations due to the oscillations of the density of
states of the ``conducting'' particles become observable at $g\sim
1$. However, as has been already mentioned, there exists another
effect, which leads to the appearance of the magnetooscillations, ---
the quantum localization. According to the scaling theory of the
quantum Hall effect \cite{pruisken},
\begin{equation}
S_{\rm QHE}(g)=2\pi g\ ,\qquad g\gg 1\ ,
\label{e31e}
\end{equation}
irrespective of any microscopic details, in particular, the value of
$\alpha$. Equations (\ref{e31d}), (\ref{e31e}) tell us that both types
of oscillations become observable at $g\sim 1$. To decide which
oscillations are stronger, one should calculate the Fourier
asymptotics (\ref{e31d}). This requires knowing the precise shape of
the function $Q(x)$. It appears that in the case of the percolation
network $Q(x)$ can be obtained only by means of a numerical
simulation. Here, we restrict ourselves to concluding that the
number of oscillations observed scales in the CF problem as $p_c\sim
k_Fd/\ln^{1/2}(k_Fd)$. Since $g\propto k_F d \exp[-(p_c/p)^2]$, where
$p=E_F/\hbar\omega_c$, the oscillations disappear extremely fast with
increasing $p$ (i.e. decreasing ${\overline B}$). These findings are
in agreement with experimental observations. The typical
number of oscillations observed in the best samples with $k_F d\sim
15$ is $p_c=7\div 9$. The oscillations start indeed to develop at $g\approx
1$, as shown in Fig.~\ref{fig9a}, where the experimental data of
Ref.~\cite{coleridge95} are represented in terms of the CF conductivity.
Finally, the damping of oscillations with decreasing ${\overline B}$
is extremely fast, so that the Dingle plot is strongly non-linear
\cite{du94,coleridge95}.

\section{ac transport}
\label{s4}

The frequency-dependent dissipative conductivity involves the Fourier
transform of the retarded velocity-velocity correlator:
$\sigma_{xx}(\omega)=e^2\rho_F\left<|v_x|^2_\omega\right>$, where
\begin{equation} 
\label{e32}
\left<|v_x|^2_\omega\right>={\rm Re}\int_0^\infty dte^{i\omega
t}\left<v_x(0)v_x(t)\right>~.  
\end{equation} 
In the quasiclassical
limit, which we consider here, the integral over $t$ is understood as
the integral along a classical trajectory, characterized by the
velocity ${\bf v}(t)$, while $\left<\,\,\right>$ denotes the averaging
over the trajectories. The average is taken over all trajectories of
electrons at the Fermi level. However, as discussed above, if the
magnetic field, either $\overline B$ or $B_0$, is strong enough, most
trajectories do not make a significant contribution to the dc
conductivity and the transport is governed by the percolation of a
small portion of their total number. In this section, we turn to the
ac conductivity. We will show that changing the frequency, one
effectively probes the motion on the percolating network on different
spatial scales. This results in a strong non-Drude frequency dispersion  
of the conductivity.

\subsection{{\it ac} conductivity at
$\alpha\sim 1$, ${\overline B}\gg B_0$: 
application to the composite fermions around $\nu=1/2$.}
\label{ss4a}

We start with the ac conductivity at $\alpha\sim 1$ in a strong
external field ${\overline B}\gg B_0$ (which is relevant to the CF
problem at sufficiently strong deviation from $\nu=1/2$, $\Delta\nu\gg
1/k_F d$). 
Our analysis in Sec.\ \ref{s3}
relies on the characterization of the electron dynamics at ${\overline
B}\gg B_0$ by three degrees of freedom with different frequency
scales: the fast cyclotron rotation, the slow drift, and the still
slower nonadiabatic diffusion across the drift trajectories. It is the
slowest degree of freedom, the nonadiabatic diffusion, that yields
unbounded (extended) paths. Let us first consider the contribution to
the velocity-velocity correlator (\ref{e32}) which comes from these
extended trajectories, $\left<v_x(0)v_x(t)\right>_{\rm
ext}$. Following the approach developed in Sec.\ \ref{s3}, we
parametrize it as a function of two variables 
\begin{equation}
\label{e33} 
\left<v_x(0)v_x(t)\right>_{\rm
ext}=v_d^2{d\over\xi}F_v({v_dt\over d},{v_dt\over L})~,
\end{equation}
where $v_d=(3/8)^{1/2}v_F R_c B_0/d{\overline B}$ is the r.m.s. drift 
velocity,  
$\xi\propto R_d^{-\nu}$ and $L\propto R_d^{-\nu-1}$ are the
characteristic size and perimeter of the elementary cell of the
percolation network, and $\nu=4/3$. The factor $d/\xi$ gives the
partial density of the percolating states (i.e., the portion of the
area $\xi\times\xi$ occupied by the trajectories of size $\xi$,
$LR_d/\xi^2\sim d/\xi$). At $d/v_d\ll t\ll L/v_d$, the scaling
function exhibits a power-law behavior $F_v(\tau,0)\sim \tau^{-x}$
reflecting the fractal dimensionality of the links of the network.
The exponent $x$ can be found by equating 
$\left<v_x(0)v_x(t)\right>_{\rm ext} t$ and the effective
diffusion coefficient $\xi^2(t)/t$, where $\xi(t)\sim
d[L(t)/d]^{\nu/(\nu+1)}$ and $L(t)\sim v_dt$, which yields
 $x=2/(\nu +1)=6/7$. The
network model we have used is justified by the condition $x<1$ (the
Harris criterion, $\nu>1$), so that the integral over $t$ in Eq.\
(\ref{e32}) at $\omega=0$ is determined by $t\sim L/v_d$. We thus have
\begin{equation} 
\label{e34} 
\left<v_x(0)v_x(t)\right>_{\rm ext}\sim
v_d^2{d\over\xi}\left( {d\over v_dt}\right)^x~,\quad {d\over v_d}\alt
t\alt {L\over v_d}~.  
\end{equation}
Now, to calculate the
frequency-dependent correction to the dc conductivity at $\omega\alt
v_d/L$, we need to know the behavior of the correlator (\ref{e33}) at
$t\gg L/v_d$. If the diffusion over the percolation network were
completely uncorrelated, $F_v$ would decay exponentially with
increasing $t$. We will argue below that in fact $F_v$ changes sign at
$t\sim L/v_d$ and falls off as a power law at larger $t$:
$$F_v(\tau,\tau')\sim -(\tau'/\tau)^x/\tau'^2\ ,\qquad \tau\gg
\tau'\gg 1.$$ 
For the correlator
(\ref{e33}) this gives 
\begin{equation} 
\label{e35}
\left<v_x(0)v_x(t)\right>_{\rm ext}\sim -{d\xi\over t^2}~,\quad t\agt
{L\over v_d}~.  
\end{equation} 
This long-time tail in the correlation
function is similar to the one found in the Lorentz gas
(noninteracting classical particles scattered by a random array of
hard discs) \cite{tails}. We will study the long-time correlations
microscopically for the realistic case of a weak long-range potential
elsewhere; here, we introduce a simple phenomenological model suitable
for the qualitative description of the percolation network.

Consider the diffusion equation with an inhomogeneous diffusion
coefficient $D({\bf r})$. The diffusion current is then given by
$D({\bf r})\nabla n({\bf r},t)$, where $n$ is the concentration of
particles. The diffusion propagator $n_{\omega {\bf q}}$
in $(\omega, {\bf q})$ space obeys the equation 
\begin{equation} 
\label{e36} 
-i\omega n_{\omega {\bf
q}}+\int {d^2{\bf q}'\over (2\pi )^2} ({\bf qq}')D_{{\bf q}-{\bf
q}'}n_{\omega {\bf q}'}=1~.  
\end{equation} 
We now write $D({\bf
r})=D+\delta D({\bf r})$, where $D$ is the mean value of $D({\bf r})$,
expand the propagator in powers of $\delta D$, and average Eq.\
(\ref{e36}) over the fluctuations with the correlator $\left<\delta
D(0)\delta D({\bf r})\right>=\Lambda D^2g(r)$, where $g(r)$ is a
dimensionless function of order unity which falls off on the scale of
$d_0$. The use of the diffusion equation implies that the correlation
radius $d_0\gg l$, where $l$ is the mean free path. To first order in
$\Lambda$ we have for the correction to the average diffusion
propagator 
\begin{equation} 
\label{e37} 
\delta n_{\omega {\bf
q}}=\Lambda D^2{\cal D}_{\omega {\bf q}}^2\int {d^2 {\bf q}'\over
(2\pi )^2}\,g_{{\bf q}-{\bf q}'}\,({\bf qq}')^2{\cal D}_{\omega {\bf
q}'}~, 
\end{equation} 
where ${\cal D}_{\omega {\bf q}}=(-i\omega
+Dq^2)^{-1}$. From Eq.\ (\ref{e37}) we deduce the $\omega$ dependent
correction to the conductivity $\delta\sigma_{xx}
(\omega)=e^2\rho\omega^2\lim_{q\to 0}q^{-2}{\rm Re}\delta n_{\omega
{\bf q}}$: 
\begin{equation} 
\label{e38} 
{\delta\sigma_{xx}
(\omega)\over \sigma_{xx}}=-{\Lambda\over 2}\int {d^2 {\bf q}\over
(2\pi )^2}\,\,g_{\bf q}\,\,{Dq^2\over -i\omega +Dq^2}~, 
\end{equation}
i.e., $\delta\sigma_{xx}(\omega)/\sigma_{xx}\simeq \Lambda
g_0|\omega|/16D$ at small $\omega$ [we drop here an $\omega$
independent term in $\delta\sigma_{xx}(\omega)$], which implies a
$t^{-2}$ long-time tail in the velocity-velocity correlator in
\mbox{Eq.\ (\ref{e32})}.

On the percolation network,
the size of the effective scatterers and the effective mean free
path are both of the order of $\xi$, i.e. $d_0\sim l\sim\xi$. 
Furthermore, strong fluctuations of the geometry of the percolating
cluster imply that $\Lambda\sim 1$. We expect that the
$1/t^2$ tail, the existence of which has been demonstrated above in the
phenomenological model with $d_0\gg l$ and $\Lambda\ll 1$, will not
disappear if we set $d_0\sim l$ and $\Lambda\sim 1$. 
 Substituting these estimates
into Eq.\ (\ref{e38}) we get 
\begin{equation} 
\label{e39}
{\delta\sigma_{xx} (\omega)\over\sigma_{xx}}\sim {|\omega|L\over
v_d}~,\quad |\omega|\alt {v_d\over L}~, 
\end{equation} 
which corresponds to Eq.\ (\ref{e35}) \cite{sign}. We recall that the
characteristic scale $L$ entering Eq.~(\ref{e39}) is given by
\begin{equation}
\label{e39a}
L\sim d\left[{me^2v_d d\over \sigma_{xx} \hbar^2}\right]^{7/3}
\propto d\exp\left({7\over 13}S_{\rm min}\right)\ .
\end{equation}

It is worth noting that the {\it classical} kinetic correlations
compete with the quantum ones and win, unless the frequency is
exponentially small. Specifically, as is well known, the quantum
localization effects in $2d$ lead to a $t^{-1}$ tail in the correlator
(\ref{e32}) and, correspondingly, to a $\ln|\omega|$ correction to the
conductivity. This quantum correction is of special interest because
of its divergence for $\omega\to 0$, in the thermodynamic limit. 
The classical correction,
proportional to $|\omega|$, does not diverge, but it is also
interesting, both theoretically and experimentally, since it is {\it
nonanalytical} in $\omega$ and is much larger than the quantum one
even at very low $\omega$. The point is that the localization
correction is a series in powers of the small parameter $1/k_Fl$,
where $k_F$ is the Fermi wave vector, while the relevant parameter for
the classical corrections is $d/l$, where $d$ is the correlation
radius of disorder. If the disorder is long-ranged ($k_Fd\gg 1$), the
classical corrections are dominant in a wide range of $\omega$. We will
discuss the classical corrections in more detail elsewhere.

Note that $\sigma_{xx}(\omega)$ behaves nonanalytically at
half-filling also in the {\it integer} quantum Hall regime. At the
integer quantum Hall transition, the frequency dependent correction 
$\delta\sigma_{xx}\sim|\omega|^{y/2}$ is related to corrections to scaling.
The leading irrelevant scaling exponent $y$
at the quantum phase transition was found numerically to be equal
to $0.38\pm 0.04$ and $0.35\pm 0.05$ in Refs.~\cite{huckestein95}
and \cite{evers98a}, respectively. In \cite{polyakov98} $y$ was argued
to be equal to $\eta$, where $\eta\simeq 0.4$
\cite{chalker88,huckestein95}  is the critical exponent of
eigenfunction correlations. However, in a
long-range random potential there exists a classical non-analytic term
$\delta\sigma_{xx}\sim|\omega|^{y_{\rm cl}/2}$ 
with $y_{\rm cl}\approx 2$ \cite{evers94}, which dominates 
$\delta\sigma_{xx}(\omega)$ in a wide range of frequencies.

Coming back to Eq.\ (\ref{e39}), we see that the frequency-dependent
correction becomes strong at $\omega\sim v_d/L$. We now turn to higher
frequencies. The scaling form (\ref{e34}) implies that the
contribution to $\sigma_{xx}(\omega)$ from the extended trajectories
behaves as $\omega^{-1+x}$ at $\omega\gg v_d/L$, i.e., it slowly
decreases as $\omega^{-1/7}$ with increasing frequency. Let us show
that the extended trajectories do not determine the conductivity any
more and the main contribution to $\sigma_{xx}(\omega)$ comes now from
localized drift trajectories. We will show that in fact
$\sigma_{xx}(\omega)$ {\it grows} with $\omega$. We neglect the weak
nonadiabatic scattering between the closed drift trajectories and
represent the velocity-velocity correlator (\ref{e32}) for the
periodic drift orbits in the form \begin{equation} \label{e40}
\left<|v_x|^2_\omega\right>_{\rm loc}=\pi\omega^2\left<
\left|\int_0^T{dt\over T}x(t)e^{i\omega t} \right|^2\sum_n\delta
\left(\omega -{2\pi n\over T}\right)\right>.  \end{equation} Here
${\bf r}(t)$ is a closed trajectory with the period $T$ and the
angular brackets denote the average over both $T$ and the shape of the
trajectory at given $T$. The average is determined by trajectories
with $T\sim \omega^{-1}$, their perimeter and size are $L_\omega\sim
v_d/\omega$ and $\xi_\omega\sim d(L_\omega/d)^{\nu/(\nu+1)}$,
respectively. The partial density of states corresponding to these
trajectories is $\sim \rho d/\xi_\omega$ [see the paragraph after Eq.\
(\ref{e33})]. The estimate for the correlator (\ref{e40}) thus reads
\begin{equation} \label{e41} \left<|v_x|^2_\omega\right>_{\rm loc}\sim
\omega^2\xi_\omega^2 \times d/\xi_\omega \times
|\omega|^{-1}~,\end{equation} where the factor $d/\xi_\omega$ stands
for the partial density of states and the last factor $\omega^{-1}$
comes from the averaging of the delta functions in Eq.\
(\ref{e40}). This yields the ac conductivity of the form
\begin{equation} \label{e42} \sigma_{xx}(\omega)\sim e^2\rho
v_dR_\omega~, \end{equation} where $R_\omega\sim
d(d/\xi_\omega)^{1/\nu}\propto |\omega|^{3/7}$ is the characteristic
width of the links of the conducting network. We get
\cite{knaebchen-note}
\begin{equation}
\label{e43} 
\sigma_{xx}(\omega)\sim \sigma_{xx}(0)\left({|\omega|
L\over v_d}\right)^{3/7}~,\quad {v_d\over L}\alt |\omega|\alt
{v_d\over d}~.  
\end{equation}

Equations (\ref{e39}),(\ref{e43}) describe the behavior of the
conductivity at $\omega\alt v_d/d$. At still higher frequencies,
$\sigma_{xx}(\omega)$ is determined by the velocity-velocity
correlations in the crossover region between the ``ballistic" drift
on the spatial scale much smaller than $d$ and the ``diffusive" motion
over the fractal network on larger scales. At the crossover, the
velocity-velocity correlator can be parameterized as
$\left<v_x(0)v_x(t)\right>=v_d^2F_v(v_dt/d)$, where $F_v(\tau)$ is a
dimensionless function of order unity [cf.\ Eq.\ (\ref{e33})]. Note
that it is not sufficient at $\omega\gg v_d/d$ to know the behavior of
the correlator at $t\sim \omega^{-1}$. Since the conductivity is
expressed in terms of the high-frequency Fourier component of
$\left<v_x(0)v_x(t)\right>$, which is an even function of $t$,
$\sigma_{xx}$ will fall off exponentially with increasing $\omega$
(until $\omega$ reaches the low-frequency wing of the cyclotron
resonance). Therefore, to get the asymptotic behavior of
$\sigma_{xx}(\omega)$, we need to know the analytical properties of
the correlator as a function of $t$, i.e., the exact shape of the
function $F_v(\tau)$, which requires a numerical simulation. We
expect, however, that the function $F_v(\tau)$ has a simple analytical
structure with singular points at ${\rm Im}\tau\sim \pm 1$, which
yields the exponential falloff of the form 
\begin{equation}
\label{e43a}
\ln\sigma_{xx}\sim -|\omega| d/v_d.
\end{equation}
This exponential decay of $\sigma_{xx}$ is limited from the side of
large frequencies by the disorder-broadened
cyclotron resonance which dominates
$\sigma_{xx}(\omega)$ at $\omega\sim\omega_c= e{\overline
B}/mc$. 

The arguments of the last paragraph concerning the exponential falloff
at large $\omega$ are also applicable to the {\it ac} conductivity in
zero (or low) ${\overline B}$, with the only substitution of the
Fermi velocity $v_F$ for the drift velocity $v_d$. In contrast to the
Drude (white-noise disorder) case, where the velocity-velocity
correlation function has a cusp at $t=0$ leading to the slow
$1/\omega^2$ decrease of $\sigma_{xx}(\omega)$, in the case 
of smooth disorder $\langle v_x(0) v_x(t)\rangle$ is an analytic
function of $t$ at $t=0$, which implies an exponential decay of 
$\sigma_{xx}(\omega)$ at $\omega\gg v_F/d$.

\begin{figure}
\centerline{\epsfxsize=75mm\epsfbox{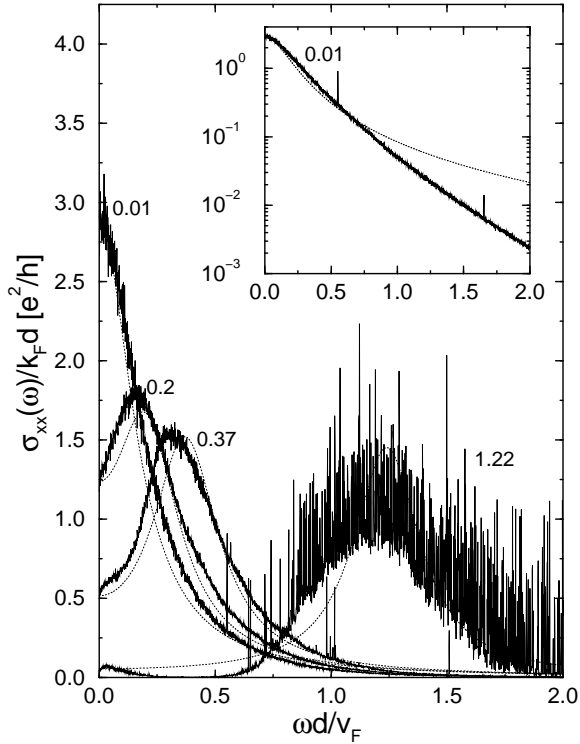}} 
\vspace{1.5cm}
\caption{{\it ac} conductivity at $\alpha=0.35$ and for different
values of the magnetic field ${\overline B}$. The number near each
curve indicates the value of the cyclotron frequency
$\omega_c=e{\overline B}/mc$ in units of $v_F/d$; to convert these
values into ${\overline B}/B_0$, one has to multiply them by
$1/\alpha$. The result of Drude theory is also shown (dotted lines).
The statistical noise in the data at $\omega_cd/v_F=1.22$ is due to
fluctuations of the local cyclotron frequency. Inset: the
low-$\overline B$ data ($\omega_cd/v_F=0.01$) on a logarithmic scale.}  
\label{fig10} 
\end{figure}

\begin{figure}
\centerline{\epsfxsize=75mm\epsfbox{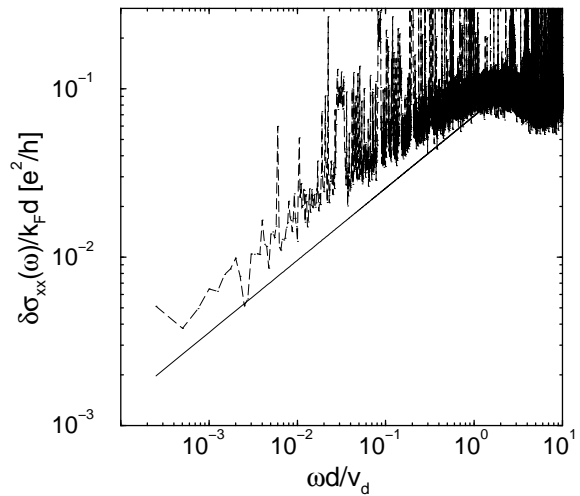}} 
\caption{Low-frequency behavior of the {\it ac} conductivity
$\delta\sigma_{xx}(\omega)=\sigma_{xx}(\omega)-\sigma_{xx}(0)$  in
strong magnetic 
field  for $\alpha=4.04$,  ${\overline B}/B_0=1.00$ (which corresponds to
$\omega_cd/v_F=4.04$). The straight line
corresponds to the theoretical prediction
$\delta\sigma_{xx}\propto|\omega|^{3/7}$, see Eq.~(\ref{e43}).
The same non-analytic behavior of $\sigma_{xx}(\omega)$
in the low-$\omega$ range in strong magnetic field
${\overline B}$ takes place for small $\alpha$ and is in particular
seen in Fig.~\ref{fig12} for $\alpha=0.2$.} 
\label{fig11} 
\end{figure}

\begin{figure}
\centerline{\epsfxsize=75mm\epsfbox{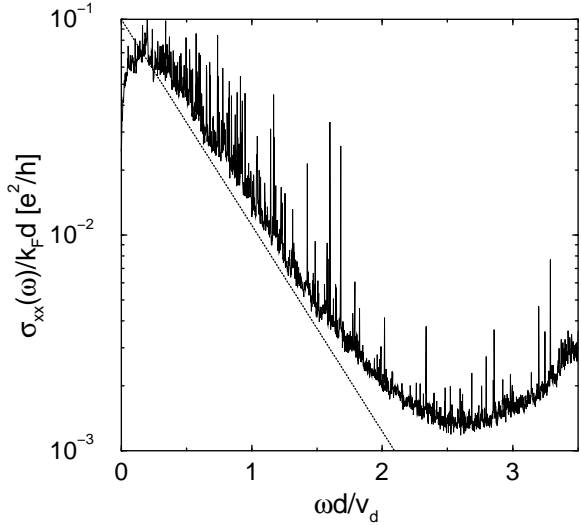}} 
\caption{{\it ac} conductivity at $\alpha=0.2$ and strong magnetic
field, ${\overline B}/B_0=4.38$ (which corresponds to
$\omega_cd/v_F=0.88$), shows  the exponential
fall-off (\ref{e43a}) in the intermediate range of frequencies. The
dotted line corresponds to  
$\ln\sigma_{xx}=-2.2|\omega|d/v_d$.} 
\label{fig12} 
\end{figure}

Figure~\ref{fig10} shows the results of the numerical calculation of the
{\it ac} conductivity for $\alpha=0.35$.  Significant
deviations from the Drude theory fit are seen, which become stronger
with increasing external magnetic field ${\overline B}$. 
For zero (or low) ${\overline B}$ the results are still
relatively close to the Drude theory, except 
in the tail (for $\omega\gg v_F/d$), where the conductivity starts to drop
exponentially (see inset), in qualitative
agreement with the theoretical expectation [Eq.~(\ref{e43a})]. 
In the intermediate fields $\overline B$ (see the curves corresponding
to $\omega_cd/v_F=0.2$ and 0.37) the non-analytic dip (\ref{e39})
around $\omega=0$ gets clearly observed. 
Finally, in a large magnetic field, the shape of the {\it ac}
conductivity is completely different: it increases non-analytically
at small $\omega$ in
agreement with Eq.~(\ref{e43}), see Fig.~\ref{fig11}, and then drops
exponentially in a higher frequency range in agreement with 
Eq.~(\ref{e43a}), see Fig.~\ref{fig12}; at still higher frequencies, the
cyclotron resonance (smeared by disorder) is observed.
It becomes difficult to resolve reliably the leading non-analytic
correction $\delta\sigma_{xx}\propto|\omega|$ at $\omega\to 0$ in high
magnetic fields, since it is shifted to the very low $\omega$ range
and is masked by the statistical noise present in the numerical data. 
Note that in Figs.~\ref{fig11} and \ref{fig12} we used the frequency
scale $v_d/d$, which is a natural scale in the regime of high magnetic
field. 
The value of $\alpha$ in Fig.~\ref{fig10} is approximately
the one appropriate for the CF system. Therefore, this
figure represents our prediction for the {\it ac} conductivity of the
CFs at half-filling (${\overline B}=0$) and away from it ($\overline
B\ne 0$).

\subsection{{\it ac} conductivity at $\alpha\gg 1$, $\overline B=0$.}
\label{ss4b}

Let us now consider the frequency dispersion of the snake-state
percolation ($\alpha\gg 1$, $\overline B=0$). Since the mechanism of
the percolation was of no importance in the derivation of Eq.\
(\ref{e39}), we get 
\begin{equation} 
\label{e44}
{\delta\sigma_{xx}(\omega)\over\sigma_{xx}}\sim {|\omega|L_s\over
v_F}~,\quad |\omega|\alt {v_F\over L_s} 
\end{equation} 
(we substituted
here $v_F$ for $v_d$ as the effective drift velocity of the snake
states and $L_s\sim d\alpha^{14/9}$ for the perimeter of the
elementary cell of the conducting network). The correction becomes
strong at $|\omega|\sim v_F/L_s$. At larger $\omega$, the conductivity
is determined by the snake states that are bounded to the closed
zero-$B$ contours of size much smaller than $\xi_s$. Specifically, the
main contribution to $\sigma_{xx}(\omega)$ now comes from the
trajectories of the length $L_\omega\sim v_F/\omega$. We estimate
their contribution at $|\omega|\agt v_F/L_s$ as
\begin{eqnarray}
\label{e45} 
\sigma_{xx}(\omega)\sim e^2\,\,{\rho
R_sL_\omega\theta_s^2\over\xi_\omega^2} \times \xi_\omega^2\omega\sim
\sigma_{xx}(0)~,\\ \nonumber {v_F\over L_s}\alt |\omega|\alt
{v_F\rho_s\over d^2}~,
\end{eqnarray} 
where $\xi_\omega\sim
d(L_\omega/d)^{\nu/(\nu+1)}$. The first factor gives the density of
states for the snake-state orbits with the periods larger than
$\omega^{-1}$, while the last factor is the effective diffusion
coefficient on the time scale of order $\omega^{-1}$. We see that
$\omega$ cancels out in this expression and $\sigma_{xx}(\omega)$
turns out to be of order $\sigma_{xx}(0)$ [\mbox{Eq.\
(\ref{e13})}]. We put $|\omega|\sim v_F\rho_s/d^2\sim
v_F/d\alpha^{2/3}$ as the upper limit for the frequency range where
Eq.\ (\ref{e45}) is valid: strictly speaking, the decrease of
$L_\omega$ at higher $|\omega|$ is accompanied with a growth of the
phase volume of the snake states that participate in the ac transport
by escaping the adiabatic traps. This means that the angle $\theta$
which should be substituted for $\theta_s$, Eq.\ (\ref{e12}), now
increases with $|\omega|$ as ${\cal L}_\omega^{-1/2}$, where ${\cal
L}_\omega\sim \ln^{1/4}(v_F/|\omega|d)$. It follows that
\begin{equation} 
\label{e46} 
\sigma_{xx}(\omega)\sim
\sigma_{xx}(0){{\cal L}\over {\cal L}_\omega}~,\quad {v_F\rho_s\over
d^2}\alt |\omega|\alt {v_F\over d}~,
\end{equation} 
i.e., the
conductivity grows with increasing $\omega$ (but extremely slowly),
until $|\omega|$ becomes of the order of $v_F/d$.

Note that, in the derivation of Eqs.\ (\ref{e45}),(\ref{e46}), we
restricted ourselves to the snake states and were not concerned about
the contribution of the drift trajectories, which requires
comment. Indeed, one might think that the arguments that led to the
power-law behavior of $\sigma_{xx}(\omega)$ in Eq.\ (\ref{e43}) could
be used here as well. However, there is a new feature that makes the
percolation at large $\alpha$ and zero $\overline B$ distinct from
that at $\alpha\sim 1$ and large $\overline B$. Namely, now there is
no characteristic drift velocity $v_d$ which is the same for all
trajectories.  Specifically, $v_d$ at large $\alpha$ depends on the
typical distance ${\cal R}$ between the drift trajectory and the
zero-$B$ contour. We parametrize $v_d({\cal R})$ as 
\begin{equation}
\label{e47} 
v_d({\cal R})=v_FF_s\left({{\cal R}\over R_s} \right)~,
\end{equation} 
where $F_s(0)\sim 1$ and $F_s(x)\sim -x^{-2}$ at $x\gg
1$. The slowing down of the drift with increasing $\cal R$ is related
to the linear growth of the magnetic field as one moves away from the
line $B=0$. Equation (\ref{e42}) tells us that
the contribution to the ac conductivity of trajectories separated by
the distance $\cal R$ from the zero-$\delta B$ lines scales as
$v_d({\cal R}){\cal R}\propto {\cal R}^{-1}$, i.e., it {\it decreases}
with $\cal R$. It follows that the drift orbits that surround the
snake-state trajectories do not contribute to $\sigma_{xx}(\omega)$
even at large $\omega$ \cite{ss/co-trans}.

Thus, the overall picture in the snake-state percolation regime is as
follows: $\sigma_{xx}(\omega)$ exhibits a narrow dip around
$\omega=0$,  increasing linearly with growing $\omega$; this increase
is saturated at $\omega\sim v_F/L_s$
where the $\omega$-dependent correction becomes strong; finally, on
the (parametrically larger) scale of  $\omega\sim v_F/d$,
$\sigma_{xx}(\omega)$ starts to fall off exponentially 
\cite{betise}. The latter
regime is similar to that at large $\overline B$ [see the paragraph
after \mbox{Eq.\ (\ref{e43})}].

\begin{figure}
\centerline{\epsfxsize=75mm\epsfbox{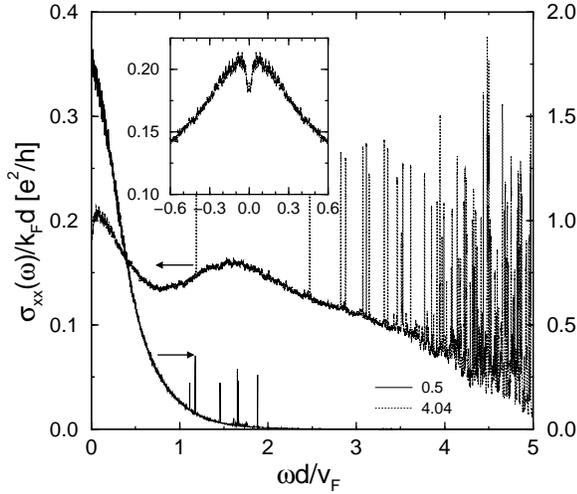}} 
\caption{{\it ac} conductivity in zero ${\overline B}$
at $\alpha=0.5$ and $\alpha=4.04$. The scaling of the $y$ axis is
different for the two curves, as indicated by the arrows.
Inset: non-analytic dip around $\omega=0$ [Eq.~(\ref{e44})]
at $\alpha=4.04$.} 
\label{fig13} 
\end{figure}

We finally comment on the case
$\alpha\sim 1$, ${\overline B}=0$ relevant to the CF problem. In this
case $L_s\sim d$, so that the range of 
applicability of Eqs.\ (\ref{e45}), (\ref{e46}) shrinks away and the
{\it ac} conductivity becomes a function of the single variable
$\omega d/v_F$: 
\begin{equation} 
\label{e48} 
\sigma_{xx}(\omega)\sim
\sigma_{xx}(0)F_\omega\left({\omega d\over v_F}\right)~,
\end{equation}
where $F_\omega (x)-F_\omega(0)\sim |x|$ at $|x|\alt 1$ and
$F_\omega(x)$ falls off exponentially at larger $x$ ($\ln F_\omega\sim
-|x|$). We have already discussed the exponential behavior at large
$\omega$ in the considered regime $\alpha\sim 1$, ${\overline B}=0$
 in the end of Sec.~\ref{ss4a}; it is clearly seen in
Figs.~\ref{fig10}, \ref{fig13}. As to the nonanalytic dip at small
$\omega$, our numerical simulations indicate that it is almost
unobservable at the values of $\alpha$ describing the CF conductivity
($\alpha\lesssim 0.7$) at ${\overline B}=0$. 
Apparently, the corresponding numerical
coefficient gets very small for such values of $\alpha$. 
With increase of either $\alpha$ (Fig.~\ref{fig13}) or ${\overline B}$
(Fig.~\ref{fig10}) the non-analytic structure gets
resolved very clearly.

In Fig.~\ref{fig13} we show the {\it ac} conductivity at relatively
large $\alpha=4.04$ in comparison with that at $\alpha=0.5$. The
$\alpha=4.04$ curve is strikingly different from the Drude behavior
and shows the features discussed above: the non-analytic increase in
the small $\omega$ region, followed by a rapid decay consistent with
the theoretically predicted exponential
falloff. At still higher frequencies a broad distribution of
cyclotron resonances in the local (random) magnetic field is observed.
A hump at $\omega d/v_F\approx 1.7$ marks the onset of this regime and
corresponds to the characteristic snake state frequency
$\omega\sim\alpha^{1/2}(v_F/d)$. 
To see the exponential falloff in a broader frequency range more
clearly, one would have to consider larger values of $\alpha$.

\section{conclusions} 
\label{s5}

We have presented a detailed analytical and numerical study of the
conductivity of a 2D fermion gas in a smooth random magnetic field, in
the whole range of the parameters $\alpha$ (strength of the random
field), ${\overline B}$ (mean magnetic field), and $\omega$
(frequency). While special emphasis has been put on the
application of our results to the composite fermion description
of the half-filled Landau level, they may be equally relevant to the
electron transport in a real random magnetic field. Below, we summarize
the main findings:

\begin{enumerate}

\item At zero magnetic field ${\overline B}$, the {\it dc} transport
has a totally different character in the regimes of weak ($\alpha\ll
1$) and strong ($\alpha\gg 1$) disorder. While in the former case
$\sigma_{xx}\propto 1/\alpha^2$ [Eq.~(\ref{e1})] is correctly given by
the Born approximation, in the latter the conductivity is
determined by the percolation of snake states yielding
$\sigma_{xx}\propto 1/\alpha^{1/2}$ (up to a negligibly weak
logarithmic correction), see Eq.~(\ref{e13}). Numerical simulations
confirm these analytical findings and allow us to find $\sigma_{xx}$ in
the crossover region $\alpha\sim 1$, see Fig.~\ref{fig7}.

\item In strong mean magnetic field ${\overline B}$ the particle
motion takes
the form of an adiabatic drift of the cyclotron orbits. A non-zero
value of $\sigma_{xx}$ in this regime is entirely due to 
exponentially weak non-adiabatic scattering processes. As a
consequence, the conductivity falls off exponentially,
$-\ln\sigma_{xx}\propto {\overline B}^2$, see Eqs.~(\ref{e22}),
(\ref{e24}), and (\ref{e25}). [At $\alpha\ll 1$, an intermediate
regime appears, $-\ln\sigma_{xx}\propto {\overline B}$,  see
Eq.~(\ref{e29}).] The numerical simulations have allowed us
to find the shape of the magnetoresistance in a wide range of
${\overline B}$ for the values of
$\alpha$ ranging from $0.2$ to $5$, see Fig.~\ref{fig9}. The
magnetoresistance $\rho_{xx}({\overline B})$ shows a sharp falloff at
large ${\overline B}$ in agreement with the analytical
results. Furthermore, at $\alpha\lesssim 0.5$ this falloff is preceded
by a positive magnetoresistance in the intermediate range of
${\overline B}$. The whole shape of $\rho_{xx}({\overline B})$ at
$\alpha\sim 0.35$ (as well as its absolute value) is surprisingly
similar to the experimental magnetoresistivity in the fractional
quantum Hall effect around $\nu=1/2$ (Fig.~\ref{fig8}).

\item In contrast to the case of a short-range random potential, the
quantum magnetooscillations of the conductivity start to develop in   
the range of the magnetic field ${\overline B}$ where the 
dimensionless conductivity $g=\sigma_{xx}/(e^2/h)$ drops to a value
$\sim 1$, in agreement with experiment (Fig.~\ref{fig9a}). These
oscillations are not related to those of the total density of states
(which are damped much more strongly), but are determined by the
oscillations of the density of states of the particles moving on the
percolating network, as well as by the quantum localization effects.

\item The {\it ac} conductivity also shows distinct features related to
the deviations of the particle kinetics from the behavior following
from the Boltzmann equation. 
While at $\alpha\ll 1$ and ${\overline B}=0$ the {\it ac}
conductivity $\sigma_{xx}(\omega)$ is relatively close to the Drude 
form (except in the tail, where it drops exponentially), at large
$\alpha$ and/or ${\overline B}$ the shape of $\sigma_{xx}(\omega)$
becomes totally different. Specifically, it shows nonanalytic behavior
in the low-frequency range, see Eqs.~(\ref{e39}), (\ref{e43}),
(\ref{e44}) and Figs.~\ref{fig10}--\ref{fig13}, 
related to the long-time tails in the velocity-velocity
correlation function and reflecting the strongly non-Boltzmann
character of the transport in the percolating regime. 
At higher frequencies, $\sigma_{xx}(\omega)$ starts to drop
exponentially (which reflects the ``ballistic'' dynamics of the snake states
or drifting orbits on short spatial scales), until it
reaches the low-frequency wing of the disorder-broadened
cyclotron resonance peak, see Eq.~(\ref{e43a}) and Fig.~\ref{fig12}. The 
{\it ac} conductivity at $\alpha=0.35$ (which is in the range of
$\alpha$ relevant to the composite fermion problem) and different
values of the mean magnetic field $\overline B$ shown in Fig.~\ref{fig10}
clearly demonstrates the anomalies which we expect to be observed in the
{\it ac} transport around $\nu=1/2$. 

\end{enumerate}

\section*{Acknowledgments} We are grateful to P.T.~Coleridge,
J.H.~Smet, H.L.~Stormer, and A.S.~Yeh for sending us the experimental data 
used in Figs.~\ref{fig8}, \ref{fig9a}, and for valuable comments
concerning the details of the experiments. We thank J.~Wilke for the
help in numerical simulations of the snake state dynamics. 
This work was supported by the SFB195 
and the Graduirtenkolleg ``Kollektive Ph\"anomene im Festk\"orper''
der Deutschen Forschungsgemeinschaft, and by the INTAS grant 
No. 97-1342.

\end{multicols}

\begin{references}
\bibitem{ln} The logarithmic factor $\cal L$ was omitted in Ref.\
\cite{mirlin98}.
\bibitem{sign} Though for the model that we consider here the sign of
the tail in the velocity-velocity correlator is negative,  
$\delta\sigma_{xx}(\omega)\propto + |\omega|$, we cannot
rule out the possibility that the sign is model dependent. This
cautionary remark is particularly relevant to the percolation models.
\bibitem{id} Note that, for the purpose of the calculation of $S_{\rm
min}$, the integration in Eq.\ (\ref{e19}) along the axis of ${\rm
Im}s$ can be done within any interval of the form $(-id+ia,id+ia)$
with arbitrary $a$. The choice $a=0$ corresponds to the condition that
the optimum fluctuation $v_d^0(s)$ is real on the real axis of
$s$.
\bibitem{ss/co-trans} In fact, we should consider here only the drift
trajectories with ${\cal R}\agt {\cal R}_c=R_s(\ln\alpha)^{1/4}$,
which are decoupled from the snake states. The drift orbits with
smaller $\cal R$ transform into the snake states with $\theta\sim 1$
and so are localized by the adiabatic ``mirrors", as discussed in
Sec.\ \ref{s2}. The estimate for ${\cal R}_c$ can be obtained by
calculating the probability for a drifting cyclotron orbit to hit a
region with an anomalously large gradient of $B({\bf r})$, which is
necessary for the adiabatic transformation of the orbit into a snake
state.
\bibitem{betise} The extremely slow growth (\ref{e46}) 
in the intermediate range of $\omega$
is hardly observable for any realistic $\alpha$.
\end{references}
\end{document}